\documentclass{aa}  

\usepackage{graphicx}
\usepackage{txfonts}
\usepackage[normalem]{ulem}
\usepackage{color}
\usepackage{bm}
\usepackage{booktabs} 
\usepackage[]{xcolor}
\usepackage{graphicx}
\usepackage{subfig}

\usepackage[colorlinks=true,allcolors=blue]{hyperref}

\begin{document}

   \title{Fundamental stellar parameters of benchmark stars \\
   from CHARA interferometry}
   \subtitle{I. Metal-poor stars}

   \author{I. Karovicova
          \inst{\ref{heidelberg},\ref{heidelberg2}}
          \and
            T. R. White\inst{\ref{sydney},\ref{aarhus}}
          \and
            T. Nordlander\inst{\ref{rsaa},\ref{astro3d}}
          \and
            L. Casagrande\inst{\ref{rsaa},\ref{astro3d}}
          \and
            M. Ireland\inst{\ref{rsaa}}
          \and
            D. Huber\inst{\ref{honolulu}}
          \and
            P. Jofr\'{e}\inst{\ref{chile}}
    }

   \institute{Zentrum f\"{u}r Astronomie der 
    Universit\"{a}t Heidelberg,Landessternwarte,
    K\"{o}nigstuhl 12, 69117 Heidelberg, Germany\\
    \email{karovicova@uni-heidelberg.de} \label{heidelberg}
    \and
    Institut f\"{u}r Theoretische Astrophysik,
    Philosophenweg 12, 69120 Heidelberg, Germany \label{heidelberg2}
    \and
    Sydney Institute for Astronomy (SIfA), School of Physics, University of
    Sydney, NSW 2006, Australia \label{sydney} 
    \and 
    Stellar Astrophysics Centre
    (SAC), Department of Physics and Astronomy, Aarhus University, Ny Munkegade
    120, DK-8000 Aarhus~C, Denmark \label{aarhus} 
    \and     
    Research School of
    Astronomy \& Astrophysics, Australian National University, Canberra, ACT
    2611, Australia \label{rsaa} 
    \and Center of Excellence for Astrophysics in
    Three Dimensions (ASTRO-3D), Australia \label{astro3d} 
    \and Institute for
    Astronomy, University of Hawai`i, 2680 Woodlawn Drive, Honolulu, HI 96822,
    USA \label{honolulu} 
    \and Universidad Diego Portales, N\'{u}cleo de
    Astronom\'{i}a, Av. Ej\'{e}rcito Libertador 441, Santiago, Chile
    \label{chile} }

   \date{Received January 28, 2020; accepted May 8, 2020}

  \abstract{Benchmark stars are crucial as validating standards for current as well as future large stellar surveys of the Milky Way. However, the number of suitable metal-poor benchmark stars is currently limited, owing to the difficulty in determining reliable effective temperatures ($T_\mathrm{eff}$) in this regime.}
   {We aim to construct a new set of metal-poor benchmark stars, based on reliable interferometric effective temperature determinations and a homogeneous analysis. 
   The aim is to reach a precision of 1\% in $T_\mathrm{eff}$, as is crucial 
   for sufficiently accurate determinations of the full set of fundamental parameters
   and abundances for the survey sources.
   }
   {We observed ten late type metal-poor dwarf and giants: HD\,2665, HD\,6755, HD\,6833, HD\,103095, HD\,122563, HD\,127243, HD\,140283, HD\,175305, HD\,221170 and HD\,224930. Only three of the ten stars (HD\,103095, HD\,122563 and HD\,140283) have previously been used as benchmark stars.
   For the observations, we used the high angular resolution optical interferometric instrument PAVO at the CHARA array.
   We modelled angular diameters using 3D limb darkening models and determined 
   effective temperatures directly from the Stefan-Boltzmann relation, with an iterative procedure to interpolate over tables of bolometric corrections. 
   Surface gravities {($\log(g)$)} were estimated from comparisons to Dartmouth stellar evolution model tracks. 
   We collected spectroscopic observations from the ELODIE and FIES
   spectrographs and estimated metallicities {($\mathrm{[Fe/H]}$)} from a 1D non-LTE abundance analysis of unblended lines of neutral and singly ionized iron.}
   {We inferred $T_\mathrm{eff}$ to better than 1\%
   for five of the stars (HD\,103095, HD\,122563, HD\,127243, HD\,140283 and
   HD\,224930). The effective temperatures of the other five stars are reliable
   to between 2-3\%; the higher uncertainty on the $T_\mathrm{eff}$ for those stars is mainly due to their having a larger uncertainty in the bolometric fluxes. 
   We also determined {$\log(g)$ and $\mathrm{[Fe/H]}$
   with median uncertainties of 0.03~dex and 0.09~dex, respectively.}
    }
   {This study presents reliable and homogeneous fundamental stellar parameters for ten metal-poor stars that can be adopted as a new set of benchmarks.
   The parameters are based on our consistent approach of combining interferometric observations, 3D limb darkening modelling and spectroscopic observations. The next paper in this series will extend this approach to dwarfs and giants in the metal-rich regime.}

   
   \authorrunning{Karovicova et al.}
   \titlerunning{Fundamental stellar parameters of benchmark stars} 
   
   \maketitle
%

\section{Introduction}

In the era of large stellar surveys, it is it essential to establish a method which reliably determine fundamental stellar parameters of the observed sources. 
Surveys as {\it Gaia} \citep{Perryman01}, APOGEE \citep{alendeprieto08}, {\it Gaia}-ESO Survey \citep{Gilmore12, Randich13}, 4MOST \citep{dejong12}, WEAVE \citep{dalton12},  GALAH \citep{desilva15} and many others are collecting extraordinary observational data. The surveys are covering millions of stars over the entire sky, allowing us to better understand stellar and Galactic structure and evolution.
However, placing stars in a detailed evolutionary context is dependent on the
accurate determination of fundamental stellar parameters of the stars such as:
effective temperature ($T_\mathrm{eff}$), surface gravity ($\log(g)$),
metallicity $[\mathrm{Fe/H}]$, and
stellar radius.

Each star observed by the survey must be analyzed by using reliable stellar models which are tested and refined against a sample of reference stars, so called benchmark stars \citep{Jofre14, Heiter15}. 
Those are stars with very well defined fundamental stellar parameters that are determined independently of
the survey. It is clear that it is crucial to establish such a set of benchmarks because robust stellar models allow the parameters of the rest of the stars in the survey to be mapped to the benchmark standard scale.

Ideally, the fundamental parameters of benchmark stars would be determined homogeneously, with both high accuracy and high precision, independently of each other, and directly (i.e.~in a model-independent way). For the fundamental stellar parameter of $T_\mathrm{eff}$, the closest realization of this ideal is with optical interferometry. 
Optical interferometry is a great technique fulfilling all these requirements
because it allows an almost independent and rigorous estimate of $T_\mathrm{eff}$. It accurately and precisely measures the angular diameter, $\theta$, and in combination with the bolometric flux, $F_\mathrm{bol}$, 
which is weakly model-dependent
via the adopted bolometric correction, the $T_\mathrm{eff}$ can be determined directly by the Stefan-Boltzmann relation:
\begin{equation}
	T_\mathrm{eff} = \left(\frac{4 F_\mathrm{bol}}{\sigma \theta^2}\right)^{1/4}.
    \label{eq:teff}
\end{equation}

Unfortunately, direct and accurate as well as precise measurement of $\theta$ using optical interferometry is limited to a relatively small number of bright stars ($V < 8$\,mag) with $\theta \gtrsim 0.3$\,mas.
Therefore, the establishment of a consistent, homogeneous sample of benchmark stars is challenging.
In an ideal case, stars in such a sample would cover a wide range of stellar parameters and abundances. Unfortunately such a set of benchmarks is currently missing. 
The stars used in the {\it Gaia}-ESO survey as benchmarks (34 {\it Gaia} FGK benchmark stars in \citealt{Jofre14} and \citealt{Heiter15})
 are collected from unrelated individual, inconsistent observations reported in the literature.
Although their effective temperatures were established directly
\citep[][]{mozurkewich03, thevenin05, wittkowski06}, the values were obtained
using different interferometric instruments and methods (Mark III, CHARA,
VINCI, etc.) and final results were obtained by applying inconsistent limb-darkening
corrections from various model atmosphere grids, resulting in
an inhomogeneous data set.

For metal-poor stars, it is particularly challenging to obtain a large set of reliable benchmark stars.
This is due to the fact the stars with low metallicities are rare and there are only a few of them that can be
observed using the 
state-of-the-art interferometric instrument
at the CHARA array. Moreover, the few observable stars with low metallicities are also rather dim
and their reliable observability is 
at the current brightness limit of the technique.
Therefore, there are currently a very few metal-poor stars 
which have had their angular diameters reliably measured, and thus their effective temperatures reliably inferred.
 To derive $T_\mathrm{eff}$ of metal-poor stars is, nevertheless,
especially crucial, as metal-poor stars hold the information about the very early 
Universe and are of a special importance for Galactic archaeology \citep{frebel15, silvaag18}.
Moreover, the demand for high accuracy, high precision stellar parameters of these stars is reflected in the need for metallicity dependent surface brightness calibration for standard candles \citep{mould19a, onozato19},
and in the need for reliable calibration of metallicity-dependent parameters for asteroseismology \citep{huber12, epstein14}.

Three very metal-poor stars HD\,103095, HD\,122563 and HD\,140283 were previously interferomerically studied
\citep{karovicova18} using the same methods described
in this paper.
These metal-poor stars are {\it Gaia} FGK benchmarks,
however, two of them HD\,103095 and HD\,140283 were not recommended as benchmarks and suggested to be removed from the sample due to $T_\mathrm{eff}$ discrepancies \citep[see][and references therein for a detailed discussion]{Heiter15}. We resolved previously reported differences between $T_\mathrm{eff}$ derived by spectroscopy, photometry and interferometry and this 
allowed the inclusion of these metal-poor stars again in
the benchmark stars sample. This thus demonstrated the robustness of our approach using the most interesting and challenging candidates.

Our overall goal is to determine fundamental stellar parameters of new and updated set of benchmark stars measured with the highest possible accuracy and precision and determined by the best available stellar models.
This paper is the first from the series of papers aiming to build a new robust sample of benchmark stars collected and analysed with a consistent approach. Here, a special consideration is given to the part of our sample covering stars with low metallicities, as they are underrepresented in benchmarks stars currently in use by large stellar surveys. In this study we present ten metal-poor stars that will be part of a larger sample of benchmarks. The consistent sample, both in observations and deriving the stellar parameters of the stars presented in this paper, will serve as validating standards for current as well 
as future large stellar surveys.

         \begin{figure}
   \centering
             \includegraphics[width=\hsize]{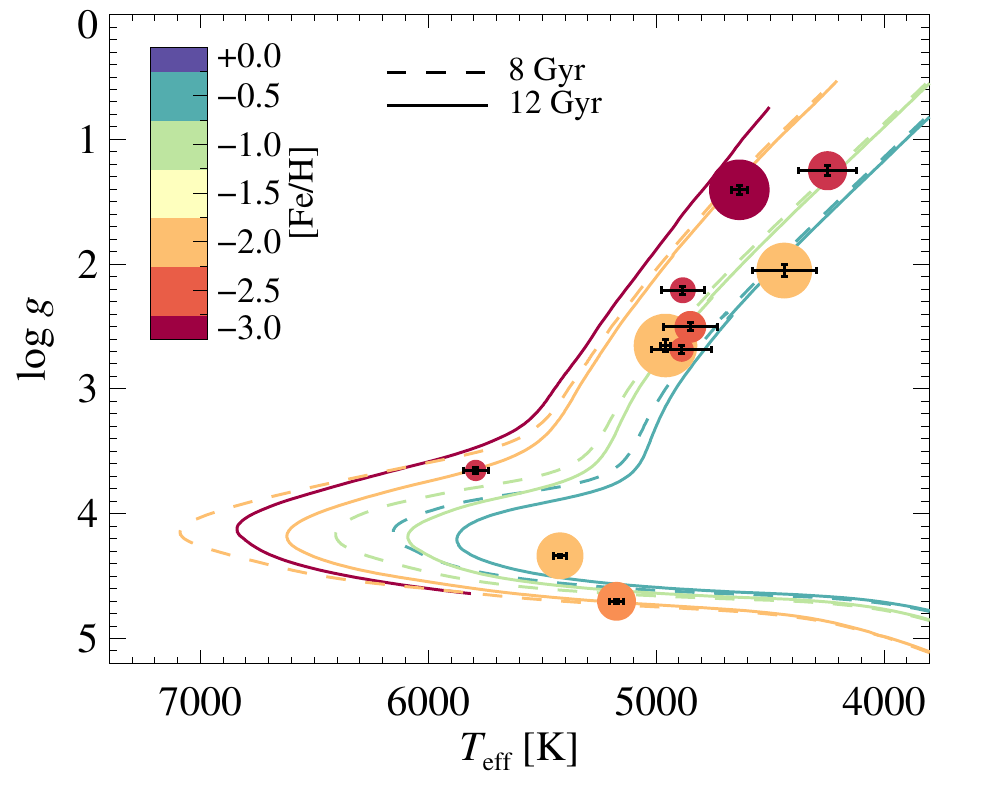} 
      \caption{Stellar parameters of our ten metal-poor stars, colour-coded by metallicity, compared to theoretical Dartmouth isochrones of different ages (linestyles) and metallicities (colours). Formal $1\sigma$ uncertainties are shown by the error bars. The symbol size is proportional to the angular diameter of the star. }
         \label{kiel}
   \end{figure}

\section{Observations}
\label{observations}
\subsection{Science targets}
\label{sciencetargets}

The ten metal-poor stars considered in this work have metallicities between $[\mathrm{Fe/H}]$\,=\,$-$0.7 and $-$2.6.  The stars are  HD\,2665, HD\,6755, HD\,6833, HD\,103095, HD\,122563, HD\,127243,
HD\,140283, HD\,175305, HD\,221170 and HD\,224930. These stars are candidates for benchmarks used for validating large stellar surveys. The sample spans the entire evolutionary range of solar mass metal poor stars as seen in Fig.~\ref{kiel}, and we list their astrometric parameters in Table~\ref{stellarparameters}.

We selected the ten stars in consultation with the {\it Gaia}-ESO spectroscopic team.
The stars have sizes and brightness such that their angular diameters can be measured reliably using the chosen interferometric instrument, and thus the $T_\mathrm{eff}$ can be inferred reliably.

Three of the stars, HD\,103095,\,HD122563, and HD\,140283, are currently used
as {\it Gaia} FGK benchmark stars \citep{Heiter15}. In the previous paper \citep{karovicova18} the reliability of the approach was demonstrated on these three stars.  They are again included here because the data reduction have been updated, and in order to present a homogeneous set of stellar parameters for all ten stars.

The other seven stars have not previously been used as benchmark stars.
HD\,175303 was discussed in an update to the {\it Gaia} FGK benchmarks
\citep{hawkins16}.  Four stars (HD\,2665, HD\,6755, HD\,6833, HD\,221170) were
suggested in the conclusion of \citet{Heiter15}, on the basis of their
inclusion in the catalogue of hydrogen line profiles from \citet{huang12}.  We
moreover added two targets (HD\,127243 and HD\,224930) with slightly higher metallicities (-0.7 dex), which according to the PASTEL catalogue \citep{soubiran10}, are thought to be typical stars and serve to complete the sample.

   \begin{table*}
  \caption{Astrometric parameters}    
  \centering   
  \begin{tabular}{l l l l l r r }      
\hline\hline   
   Star         & Right ascension &    Declination &    m$_V$ &    m$_R$ & $E(B-V)$&   $\pi$ $^a$      \\ 
   & & & (mag) & (mag) && (mas)   \\
     \hline    
    HD\,2665   &  00 30 45.447 & $+$57 03 53.627 & 7.72 & 7.74  &   0.049 $\pm$ 0.02	 & 3.714 $\pm$ 0.036    \\ 
    HD\,6755   &  01 09 43.060 & $+$61 32 50.293 & 7.68 & 7.30  &   0.010 $\pm$ 0.01	 & 5.969 $\pm$ 0.049    \\
    HD\,6833   &  01 09 52.265 & $+$54 44 20.273 & 6.74 & 6.77  &   0.047 $\pm$ 0.02	 & 4.711 $\pm$ 0.048    \\
    HD\,103095 &  11 52 58.768 & $+$37 43 07.240 & 6.45 & 5.80  &   0     $\pm$ 0	     & 108.955 $\pm$ 0.049   \\
    HD\,122563 &  14 02 31.846 & $+$09 41 09.943 & 6.19 & 5.37  &   0.003 $\pm$ 0.01	 & 3.440 $\pm$ 0.063   \\
    HD\,127243 & 14 28 37.813 & $+$49 50 41.461& 5.59 & 5.10  &   0     $\pm$ 0        & 10.390 $\pm$ 0.069  \\
    HD\,140283 &  15 43 03.097 & $-$10 56 00.596 & 7.12 & 6.63  &   0     $\pm$ 0	     & 16.144 $\pm$ 0.072   \\
    HD\,175305 &  18 47 06.442 & $+$74 43 31.448 & 7.18 & 6.52  &   0.010 $\pm$ 0.01	 & 6.349 $\pm$ 0.025   \\
    HD\,221170 &  23 29 28.809 & $+$30 25 57.847 & 7.66 & 7.69  &   0.061 $\pm$ 0.02	 & 1.837 $\pm$ 0.059   \\
    HD\,224930 &  00 02 10.341 & $+$27 04 54.477 & 5.75 & 5.16  &   0     $\pm$ 0        & 79.070 $\pm$ 0.560  \\
  \hline                                 
\end{tabular}
\tablefoot{\tablefoottext{a}{{\it Gaia} Data release 2 
}}
\label{stellarparameters}
  \end{table*}

\begin{table*}[t!]
\caption{Interferometric observations - metal-poor stars}            
\label{table:2}      
\centering                          
\begin{tabular}{l l l c c l}      
\hline\hline            
Science target & UT date & Telescope & B (m) ) & \# of obs. & Calibrator stars\\               
\hline     
   HD 2665   & 2016 Aug 11  & E1W1   & 313.57 & 3  & HD 584, HD 3519 \\ 
             & 2016 Aug 13  & E1W2   & 221.85 & 2  & HD 584, HD 3519 \\
             & 2016 Aug 17  & E2W1   & 251.34 & 1  & HD 584, HD 3519 \\   
             & 2016 Oct 7   & E2W1   & 251.34 & 5  & HD 584, HD 3519 \\    
   HD 6755   & 2016 Aug 11  & E1W1   & 313.57 & 2  & HD 3519, HD 9878 \\         
             & 2016 Aug 17  & E2W1   & 251.34 & 3  & HD 3519, HD 9878 \\  
             & 2016 Oct 7   & E2W1   & 251.34 & 6  & HD 3519, HD 9878 \\   
   HD 6833   & 2009 Jul 17  & W1W2   & 107.93 & 2  & HD 6028 \\
             & 2009 Jul 21  & S2W2   & 177.45 & 2  & HD 6676 \\
             & 2015 Sep 25  & E2W2   & 156.26 & 3  & HD 3519, HD 3802, HD 7804 \\     
   HD 103095$^a$ & 2015 May 2   & E2W2   & 156.26 & 3  & HD 99002, HD 103288 \\
             & 2017 Mar 3   & E2W2   & 156.26 & 3  & HD 99002, HD 107053 \\
             &              & E2W1   & 251.34 & 2  & HD 99002, HD 107053 \\          
             & 2017 Mar 4   & E1W2   & 221.85 & 3  & HD 99002, HD 103288, HD 107053  \\      
   HD 122563$^a$ & 2017 Mar 3   & E2W2   & 156.26 & 3  & HD 120448, HD 122365, HD 128481 \\
             & 2017 June 9  & E2W2   & 156.26 & 2  & HD 121996, HD 128481 \\
             & 2017 June 10 & E2W2   & 156.26 & 2  & HD 120934   \\
   HD 127243 & 2015 Apr 5   & W1W2   & 107.93 & 3  & HD 122866, HD 125349, HD 128184 \\     
             & 2015 Jul 27  & E2W2   & 156.26 & 2  & HD 
10
 128998, HD 133962, HD 140728 \\ 
   HD 140283$^a$ & 2014 Apr 8   & E1W1   & 313.57 & 4  & HD 139909, HD 143259, HD 146214\\ 
             & 2015 Apr 4   & S1W1   & 278.50 & 2  & HD 139909, HD 143259\\
             & 2017 June 16 & E1W1   & 313.57 & 4  & HD 128481, HD 143259\\             
   HD 175305 & 2015 Jul 28  & E2W2   & 156.26 & 3  & HD 157774, HD 169027 \\   
             & 2015 Sep 21  & E1W2   & 221.85 & 4  & HD 146929, HD 157774, HD 169027 \\          
             & 2015 Sep 23  & E1W2   & 221.85 & 2  & HD 146929, HD 169027 \\           
             & 2015 Sep 24  & E2W2   & 156.26 & 4  & HD 169027, HD 178738, HD 197637 \\          
   HD 221170 & 2009 Jul 20  & S2W2   & 177.45 & 3  & HD 221491 \\
             & 2015 Sep 8   & E1W2   & 221.85 & 1  & HD 220599 \\
             & 2016 Aug 10  & E2W2   & 156.26 & 3  & HD 220599, HD 221491 \\
             & 2016 Aug 13  & E1W2   & 221.85 & 3  & HD 220599, HD 221491 \\
             & 2016 Oct 7   & E2W1   & 251.34 & 2  & HD 220599, HD 221491 \\   
   HD 224930 & 2015 Aug 6   & S2W2   & 177.45 & 3  & HD 1439 \\           
             & 2015 Aug 7   & E2W2   & 156.26 & 3  & HD 1439, HD 1606 \\   
   \hline                                 
\end{tabular}
    \tablefoot{\tablefoottext{a}{The data of the three stars presented in the previous study \citep{karovicova18} are repeated here for completeness.}}
\end{table*}

\subsection{Interferometric observations and data reduction}
\label{observations_interferometry}
We observed the stars
using the interferometric instrument PAVO \citep{Ireland2008}.
The instrument is located at the CHARA array at Mt. Wilson
Observatory, California \citep{tenBrummelaar05}.
The PAVO instrument is operating in optical wavelengths
between $\sim$\,600--900\,nm and it is a pupil-plane beam combiner. The PAVO instrument is limited to observations of targets with magnitudes of m$_R$\,$\sim$\,7.5. 
In the case of ideal weather conditions, it is possible to observe targets down to m$_R$=8, with recent improvements due to adaptive optics \citep{che14}.
The CHARA array offers the longest available baselines in the optical wavelengths worldwide. The stars were observed using baselines between 107.9\,m and 313.6\,m. We collected the observations between 2009 Jul 17 and 2016 Oct 7.
Table~\ref{table:2} summarizes our dates 
of observations, telescope configuration and the projected baselines B.

The data were reduced with the PAVO reduction software. The PAVO data reduction software has been well-tested and used in multiple studies \citep{Bazot11,Derekas11,huber12,Maestro13}.
In order to monitor the interferometric transfer function, a set of calibrating stars were observed. These calibrating stars were selected from a catalogue of CHARA calibrators and from the Hipparcos catalogue \citep{hipparcos}. According to the location and sizes of an observed target we selected unresolved or closely unresolved sources,
located close on the sky to the science target. The calibrating stars were observed immediately before as well as after the science target.
We determined the angular diameters of the calibrators 
using the $V-K$ relation of \citet{boyajian14} and corrected for limb-darkening to determine the uniform disc diameter in $R$ band.
The $V$-band magnitudes were selected from the Tycho-2 catalogue \citep{Hoeg2000} and converted into the Johnson system using the calibration by \citet{bessell00}. The $K$-band magnitudes were selected from the Two Micron All Sky Survey \citep[2MASS;][]{Skrutskie2006}. The reddening was estimated from the dust map of \citet{green15} and the reddening law of \citet{odonnell94} was applied.
 We set the relative uncertainty on calibrator diameters to 5\%
 \citep{boyajian14}. The uncertainty is set in a way that it covers the
 uncertainty on the calibrator diameters as well as the uncertainty on the
 reddening. We also set the absolute uncertainty on the wavelength 
 scale to 5\,nm.
 We checked the literature for each calibrator to ensure they were not known
 binaries. According to Gaia\,DR2, both the proper motion anomaly \citep{kervella19} and the 
$\rm{\tt phot\_bp\_rp\_excess\_factor}$
 \citep{evans18} suggest that none of our calibrators has a companion that
 is large enough to affect our interferometric measurements or estimated calibrator sizes.
  We note that for the smallest science targets, such as HD\,2665 and HD\,6755, we have endeavored to choose the smallest calibrators that were practical, 
which in these cases were $<$\,0.15 mas.
For all the calibrating stars, their spectral type, magnitude in the $V$ and $R$ band, their expected angular diameter
and the corresponding science targets are summarized in Table~\ref{table:3}.

   \begin{table}[t!]
\caption{Calibrator stars used for interferometric observations - metal-poor stars}            
\label{table:3}      
\centering                          
\begin{tabular}{l l c c c c l}      
\hline\hline            
Calibrator & Spectral & m$_V$ & m$_K$  & $E(B-V)$     & UD  \\  
           & type         &       &     & (mag)  & (mas)         \\ 
\hline    
 HD 584     & B8III & 6.72 & 6.97 & 0.113  &  0.126 \\  
 HD 1439    & A0IV  & 5.88 & 5.86 & 0.042  &  0.221 \\  
 HD 1606    & B7V   & 5.87 & 6.23 & 0.050  &  0.177 \\ 
 HD 3519    & A0    & 6.72 & 6.74 & 0.093  &  0.145 \\ 
 HD 3802    & A0    & 6.73 & 6.57 & 0.008  &  0.164 \\ 
 HD 6028    & A3V   & 6.47 & 6.01 & 0.023  &  0.221 \\    
 HD 6676    & B8V   & 5.77 & 5.75 & 0.049  &  0.233 \\  
 HD 7804    & A1V   & 5.14 & 4.92 & 0.008  &  0.353 \\      
 HD 9878    & B7V   & 6.71 & 6.70 & 0.185  &  0.145 \\   
 HD 99002   & F0    & 6.93 & 6.28 & 0.008  &  0.201  \\  
 HD 103288  & F0    & 7.00 & 6.22 & 0.006  &  0.211 \\
 HD 103928  & A9V   & 6.42 & 5.60 & 0.002  &  0.282   \\
 HD 107053  & A5V   & 6.68 & 6.02 & 0.004  &  0.226  \\   
 HD 120448  & A0    & 6.78 & 6.52 & 0.017  &  0.169 \\  
 HD 120934  & A1V   & 6.10 & 5.96 & 0.007  &  0.216  \\  
 HD 121996  & A0Vs  & 5.76 & 5.70 & 0.029  &  0.238  \\
 HD 122365  & A2V   & 5.98 & 5.70 & 0.007  &  0.248\\
 HD 122866  & A2V   & 6.15 & 6.11 & 0.005  &  0.199 \\   
 HD 125349  & A2IV  & 6.20 & 5.98 & 0.002  &  0.217 \\  
 HD 128184  & A0    & 6.51 & 6.29 & 0.009  &  0.188 \\    
 HD 128481  & A0    & 6.98 & 6.79 & 0.007  &  0.149 \\   
 HD 128998  & A1V   & 5.82 & 5.76 & 0.009  &  0.235 \\    
 HD 133962  & A0V   & 5.58 & 5.61 & 0.003  &  0.249 \\  
 HD 139909  & B9.5V & 6.86 & 6.54 & 0.110  &  0.165 \\   
 HD 140728  & A0V   & 5.48 & 5.56 & 0.008  &  0.253 \\  
 HD 143259  & B9V   & 6.64 & 6.28 & 0.107  &  0.187 \\
 HD 146214  & A1V   & 7.49 & 7.10 & 0.012  &  0.132 \\
 HD 146926  & B8V   & 5.48 & 5.70 & 0.014  &  0.233 \\   
 HD 157774  & A0    & 7.13 & 7.01 & 0.011  &  0.133 \\    
 HD 169027  & A0    & 6.79 & 6.95 & 0.011  &  0.132 \\   
 HD 178738  & A0    & 6.89 & 6.85 & 0.036  &  0.141 \\   
 HD 197637  & B3    & 6.94 & 7.35 & 0.107  &  0.104 \\   
 HD 220599  & B9III & 5.55 & 5.72 & 0.010  &  0.232 \\   
 HD 221491  & B8V   & 6.64 & 6.75 & 0.034  &  0.145 \\   
  \hline                                 
\end{tabular}
\end{table}

\begin{table*}
\caption{Angular diameters and linear limb-darkening coefficients.}   \label{table:4}      
\centering                          
\begin{tabular}{llll}      
\hline\hline            
Star  & $\theta_\mathrm{UD}$ (mas) & \multicolumn{2}{l}{Linear limb darkening$^a$}   \\   
     &                              &    $u$    &    $\theta_\mathrm{LD}$ (mas)   \\
\hline    
    HD\,2665   & 0.377 $\pm$ 0.004 & 0.561 $\pm$ 0.009 & 0.397 $\pm$ 0.003\\
    HD\,6755   & 0.354 $\pm$ 0.004 & 0.575 $\pm$ 0.014 & 0.375 $\pm$ 0.004\\
    HD\,6833   & 0.804 $\pm$ 0.009 & 0.674 $\pm$ 0.011 & 0.862 $\pm$ 0.009 \\
    HD\,103095 & 0.565 $\pm$ 0.004 & 0.565 $\pm$ 0.016 & 0.597 $\pm$ 0.005  \\
    HD\,122563 & 0.861 $\pm$ 0.010 & 0.568 $\pm$ 0.009 & 0.907 $\pm$ 0.011 \\
    HD\,127243 & 0.922 $\pm$ 0.006 & 0.621 $\pm$ 0.013 & 0.983 $\pm$ 0.008 \\
    HD\,140283 & 0.311 $\pm$ 0.005 & 0.510 $\pm$ 0.003 & 0.326 $\pm$ 0.006 \\
    HD\,175305 & 0.461 $\pm$ 0.006 & 0.590 $\pm$ 0.014 & 0.487 $\pm$ 0.006 \\
    HD\,221170 & 0.563 $\pm$ 0.005 & 0.632 $\pm$ 0.014 & 0.599 $\pm$ 0.006  \\
    HD\,224930 & 0.680 $\pm$ 0.007 & 0.566 $\pm$ 0.014 & 0.720 $\pm$ 0.007 \\
\hline                                 
\end{tabular}
\tablefoot{\tablefoottext{a}{Limb-darkening coefficients derived from the grid of \citet{claret11}; see text for details.}}
\label{angulardiam}
\end{table*}

   \begin{table*}
   \caption{Observed ($\varTheta_{LD}$) and derived ($F_\mathrm{bol}$, $M$, $L$, $R$) stellar parameters} 
   \centering   
  \begin{tabular}{l c c c c c}      
\hline\hline   
 
 Star &  $F_\mathrm{bol}$  &     $\varTheta_{LD}$&      Mass (M$_\odot$)    &   $L$ (L$_\odot$) &    $R$ (R$_\odot$)\\
    & (erg s$^{-1}$cm$^{-2}$10$^{-8}$)&  (mas) & &  \\ 
  \hline        
   HD\,2665     & 2.95 $\pm$ 0.22   & 0.395 $\pm$ 0.004  & 0.77 $\pm$ 0.05 & 66.4 $\pm$ 5.2  & 11.43 $\pm$ 0.16    \\
   HD\,6755     & 2.59 $\pm$ 0.27   & 0.369 $\pm$ 0.004  & 0.78 $\pm$ 0.05 & 22.7 $\pm$ 2.4  & 6.648 $\pm$ 0.090     \\
   HD\,6833     & 9.4 $\pm$ 1.2   & 0.852 $\pm$ 0.008  & 1.00 $\pm$ 0.15 & 152.6 $\pm$ 5.8 &  19.45 $\pm$ 0.27   \\
   HD\,103095   & 8.41 $\pm$ 0.18   & 0.593 $\pm$ 0.004  & 0.63 $\pm$ 0.02 & 0.221 $\pm$ 0.005   & 0.586 $\pm$ 0.004     \\
   HD\,122563   & 13.14 $\pm$ 0.22  & 0.925 $\pm$ 0.011  & 0.77 $\pm$ 0.05 & 339 $\pm$ 13& 28.86 $\pm$ 0.63    \\
   HD\,127243   & 18.99 $\pm$ 0.18  & 0.971 $\pm$ 0.007  & 1.46 $\pm$ 0.15 & 54.97 $\pm$ 0.90  & 10.045 $\pm$ 0.098  \\
   HD\,140283   & 3.955 $\pm$ 0.029   & 0.325 $\pm$ 0.006  & 0.77 $\pm$ 0.03 & 4.766 $\pm$ 0.055   &  2.167 $\pm$ 0.041    \\
   HD\,175305   & 4.33 $\pm$ 0.41   & 0.484 $\pm$ 0.006  & 0.78 $\pm$ 0.05 & 33.5 $\pm$ 3.2  &  8.20 $\pm$ 0.11    \\
   HD\,221170   & 3.85 $\pm$ 0.46   & 0.596 $\pm$ 0.005  & 0.79 $\pm$ 0.05 & 3567 $\pm$ 48&  34.86 $\pm$ 1.16   \\
   HD\,224930   & 14.76 $\pm$ 0.10  & 0.716 $\pm$ 0.007  & 0.75 $\pm$ 0.01 & 0.741 $\pm$ 0.012    &  0.973 $\pm$ 0.012    \\
  \hline  
\end{tabular}
\label{tab:observed}
  \end{table*}

      \begin{figure}
   \centering
          \includegraphics[width=\hsize]{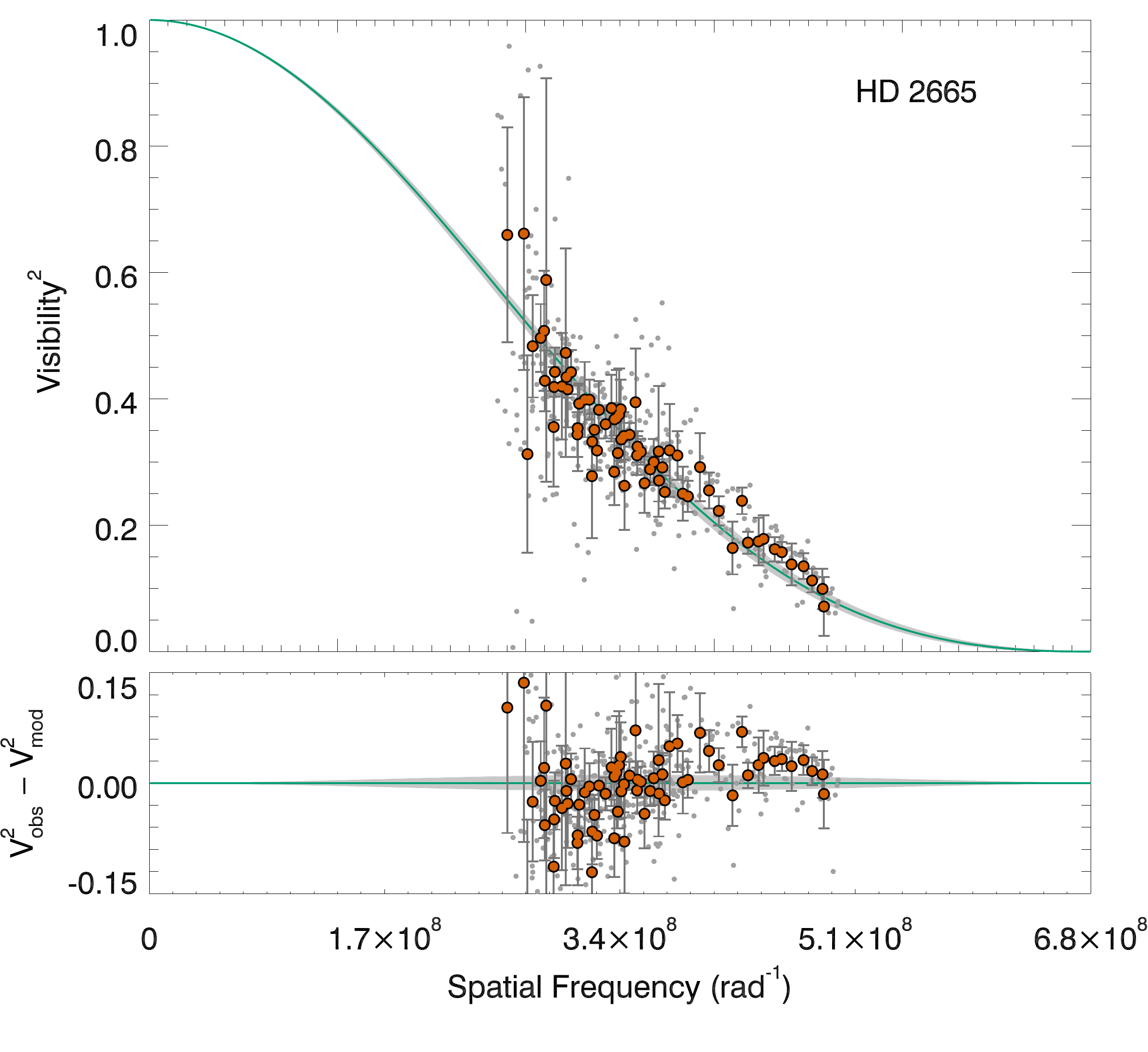}
          \includegraphics[width=\hsize]{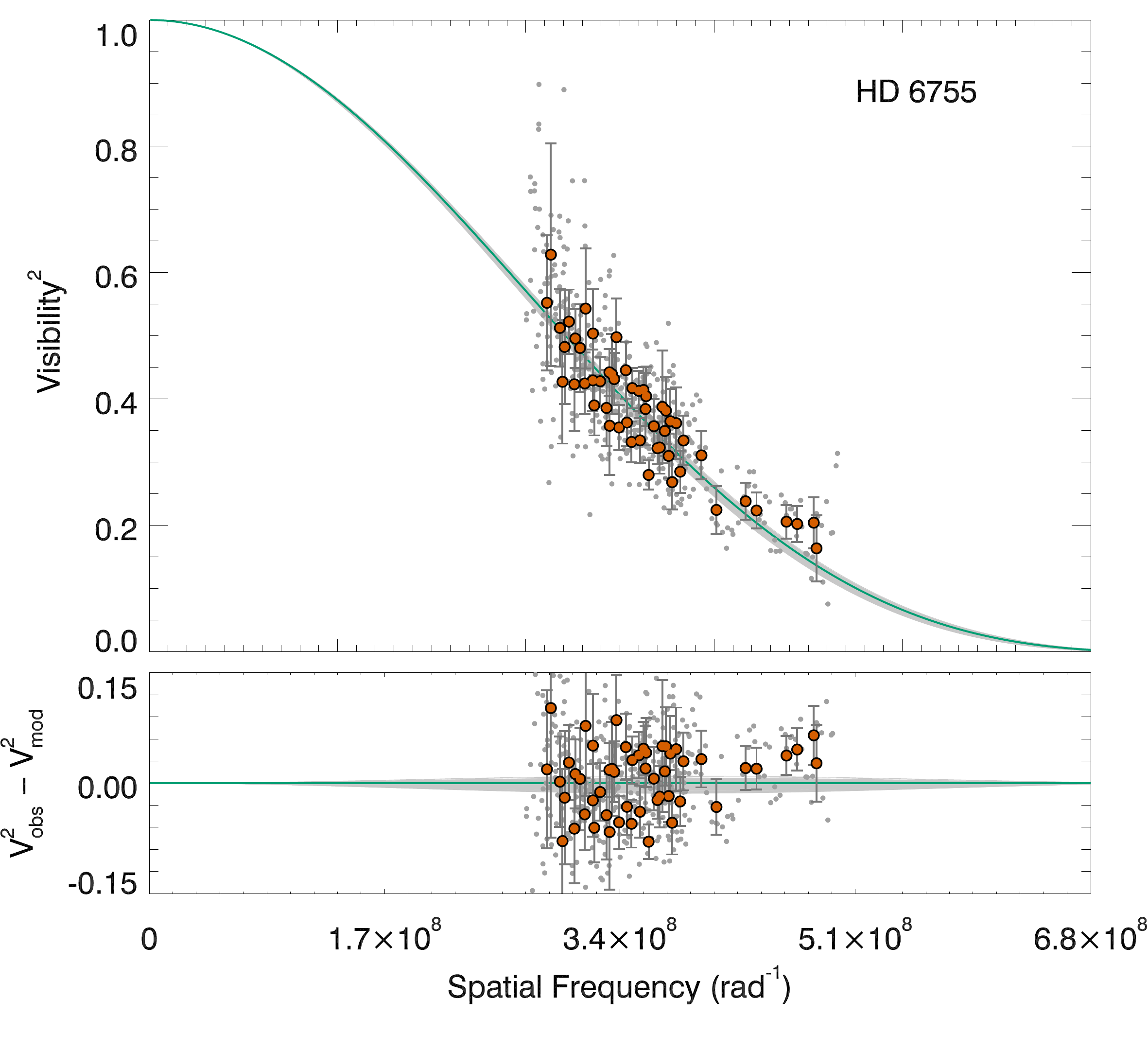}
      \caption{Squared visibility versus spatial frequency for HD\,2665 and HD\,6755. The HD number is noted in the right upper corner in the each plot.
      The error bars have been scaled to the reduced $\chi^{2}$. For HD\,2665 the reduced $\chi^{2}$\,=\,1.6 and for HD\,6755 $\chi^{2}$\,=\,1.7.
      The grey dots are the individual PAVO measurements in each wavelength channel.
      For clarity, we show weighted averages of the PAVO measurements 
      as red circles. The green line shows the fitted limb-darkened model to the PAVO data, with the light grey-shaded region indicating the 1-$\sigma$ uncertainties. The lower
      panel shows the residuals from the fit.}
      \label{figures1}
   \end{figure}

      \begin{figure}
   \centering
          \includegraphics[width=\hsize]{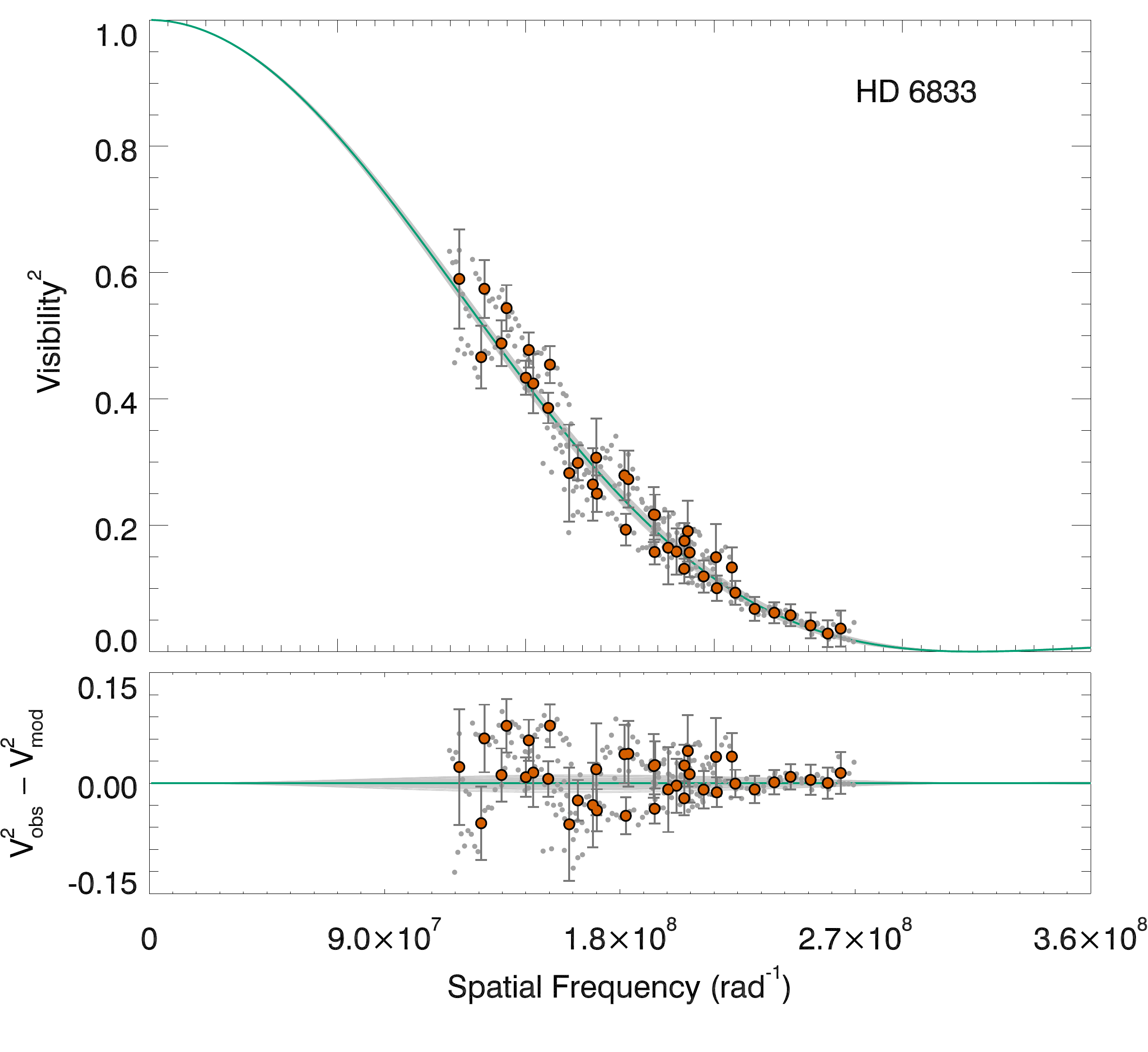}
          \includegraphics[width=\hsize]{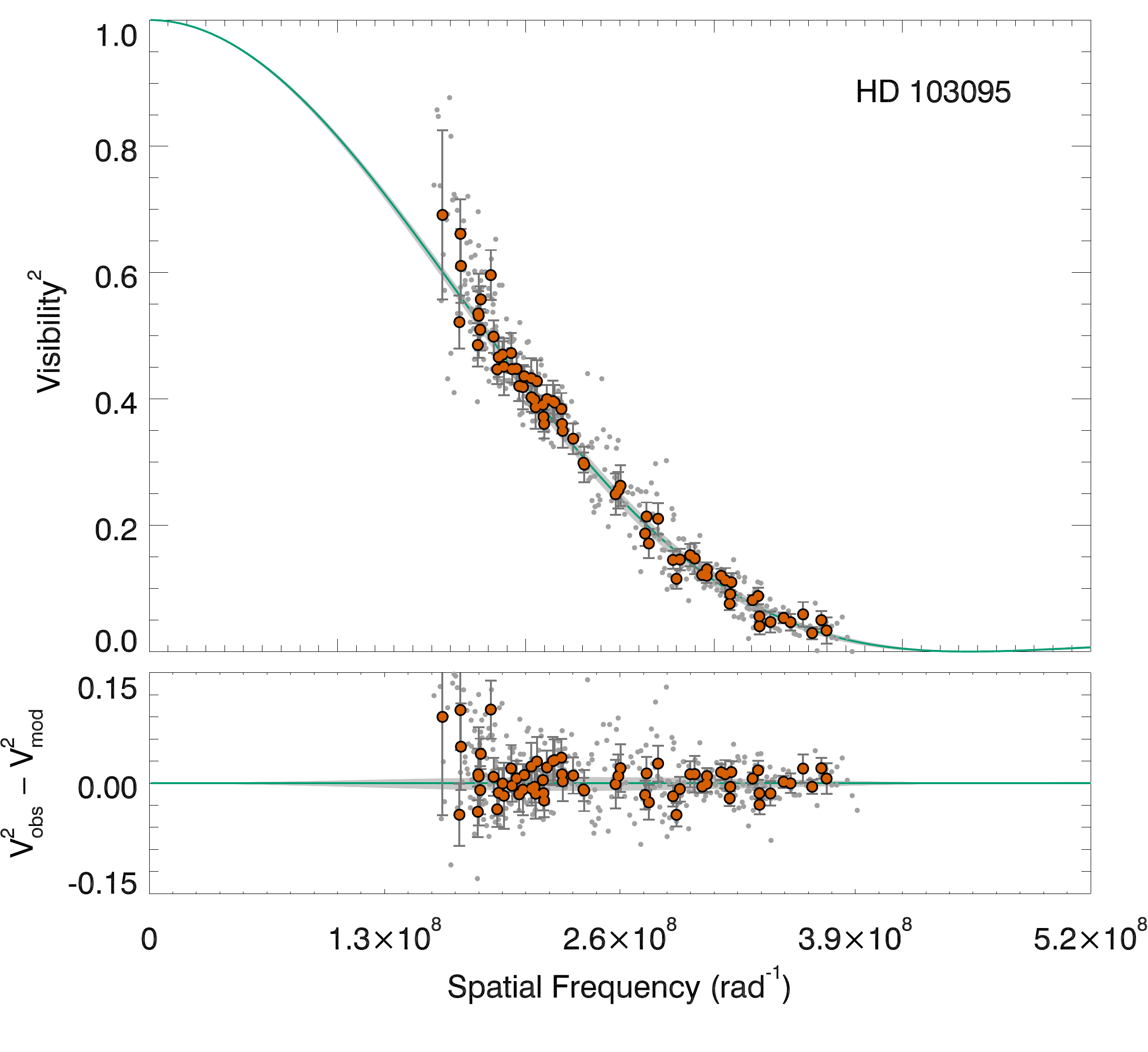}
            \caption{Squared visibility versus spatial frequency for HD\,6833 and HD\,103095.
             The lower
      panel shows the residuals from the fit.  
            The error bars have been scaled to the reduced $\chi^{2}$. For HD\,6833 the reduced $\chi^{2}$\,=\,7.0 and for HD\,103095 $\chi^{2}$\,=\,1.1.
       All lines and symbols are the same as for Fig.\ref{figures1}.        
}
         \label{figures2}
   \end{figure}

            \begin{figure}
   \centering
                \includegraphics[width=\hsize]{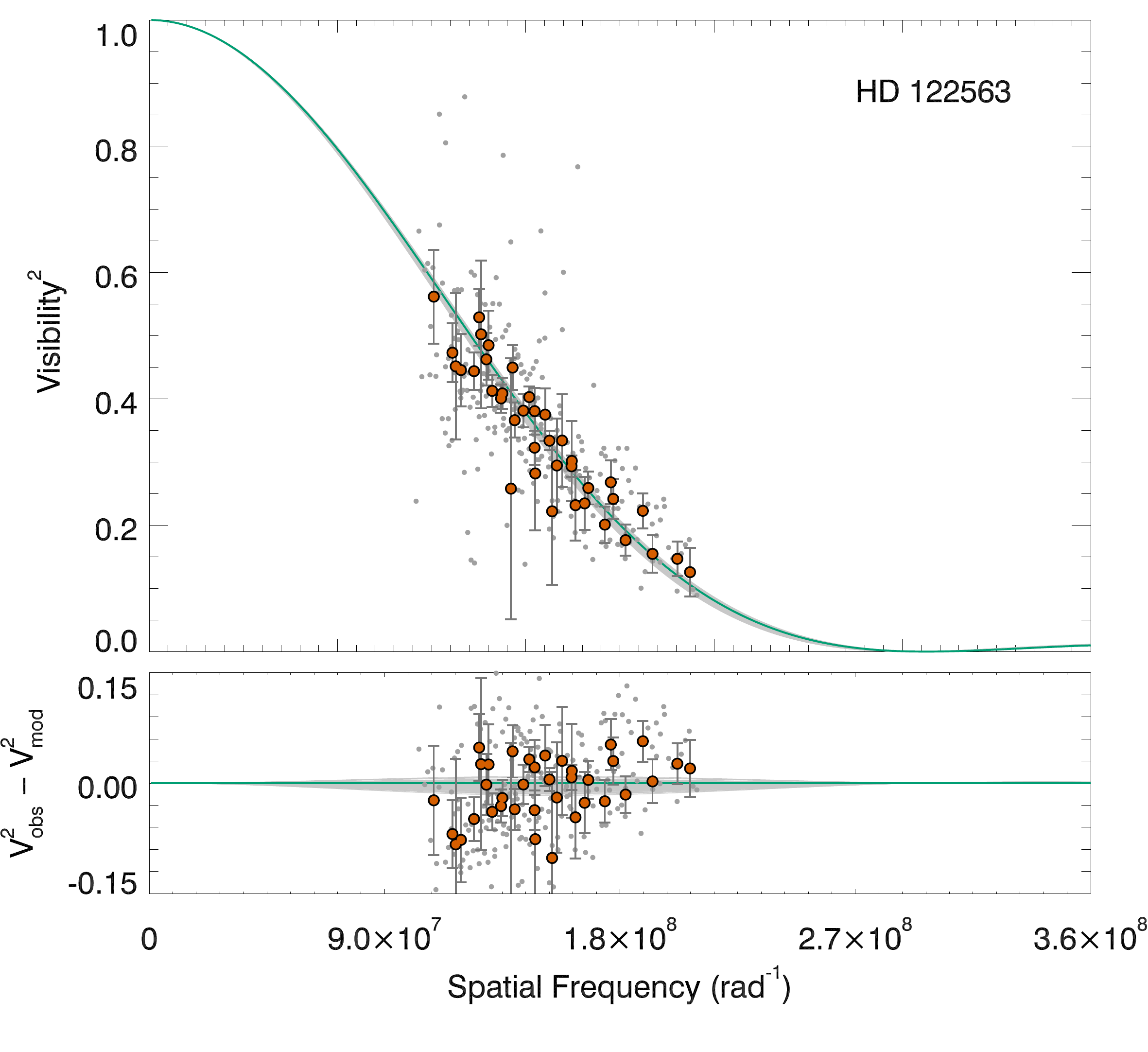}
                \includegraphics[width=\hsize]{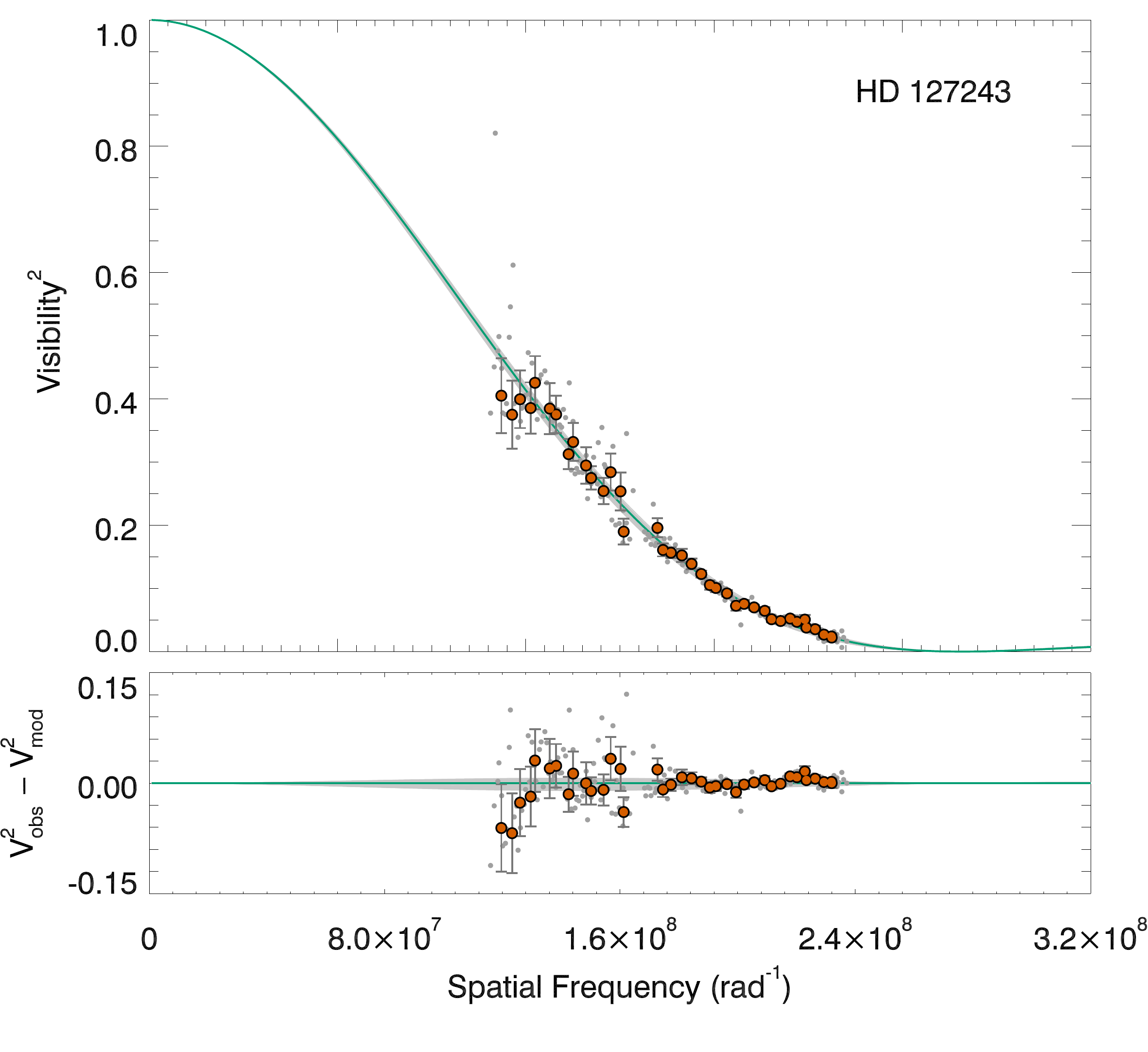}
              \caption{Squared visibility versus spatial frequency for HD\,122563 and HD\,127243.
                         The lower
      panel shows the residuals from the fit.  
       The error bars have been scaled to the reduced $\chi^{2}$. For HD\,122563 the reduced $\chi^{2}$\,=\,3.0 and for HD\,127243 $\chi^{2}$\,=\,1.6.     
       All lines and symbols are the same as for Fig.\ref{figures1}.
      }
         \label{figures3}
   \end{figure}

            \begin{figure}
   \centering
 
                \includegraphics[width=\hsize]{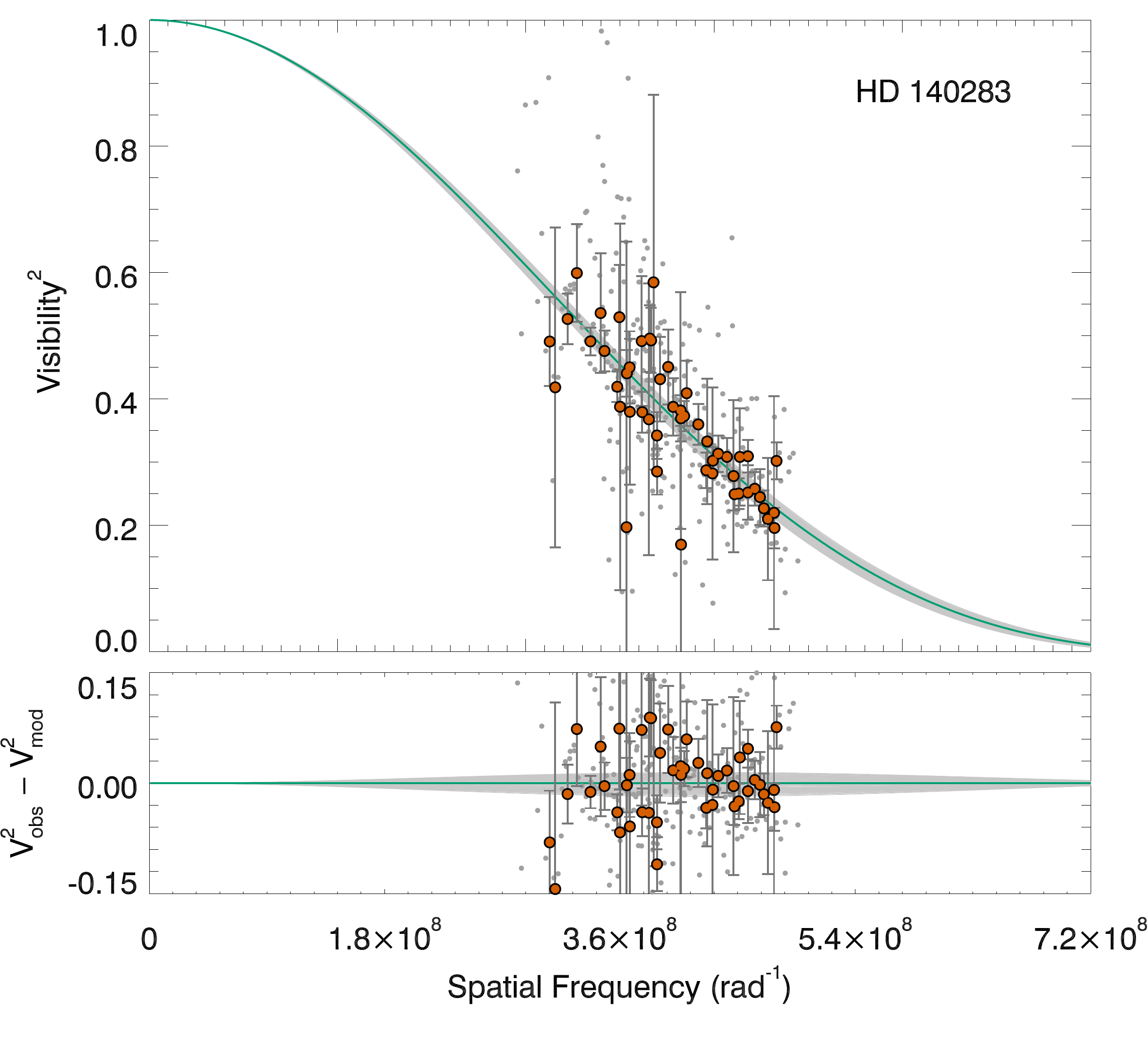}
                \includegraphics[width=\hsize]{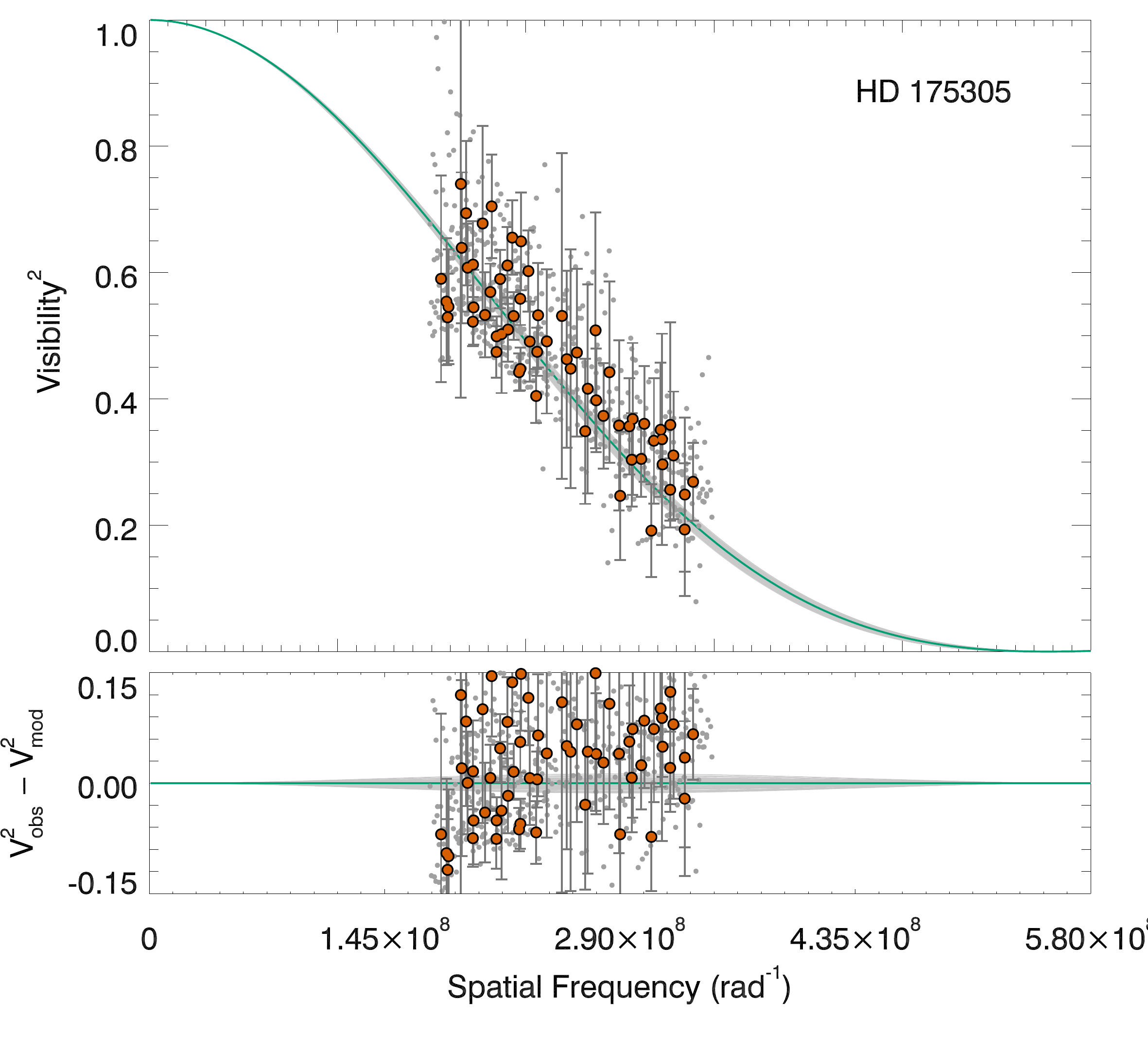}
      
            \caption{Squared visibility versus spatial frequency for HD\,140283 and HD\,175305.
                         The lower
      panel shows the residuals from the fit.      
            The error bars have been scaled to the reduced $\chi^{2}$. For HD\,140283 the reduced $\chi^{2}$\,=\,1.4 and for HD\,175305 $\chi^{2}$\,=\,3.7.
       All lines and symbols are the same as for Fig.\ref{figures1}.
     }
         \label{figures4}
   \end{figure}

         \begin{figure}
   \centering
 
             \includegraphics[width=\hsize]{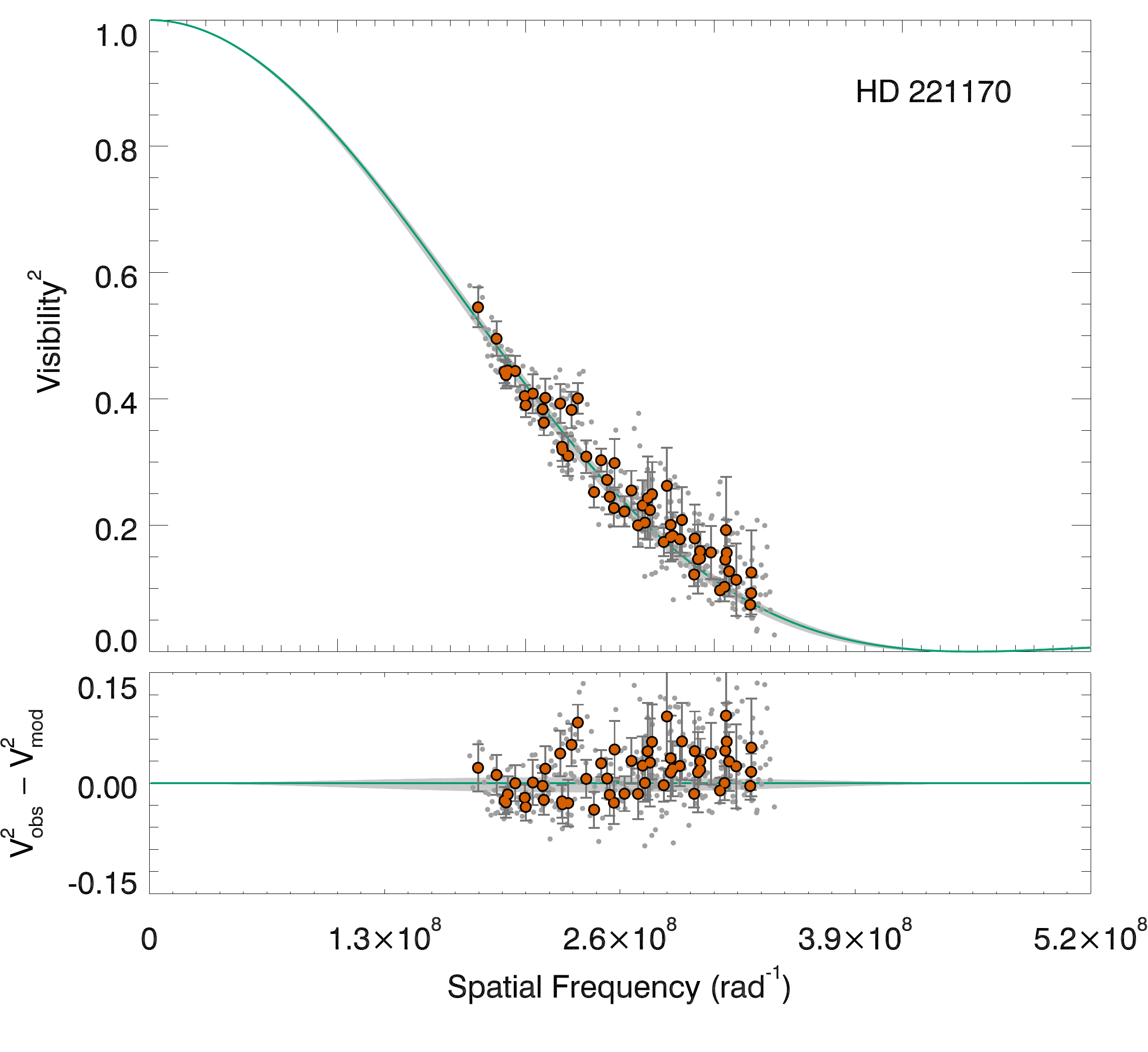} 
             \includegraphics[width=\hsize]{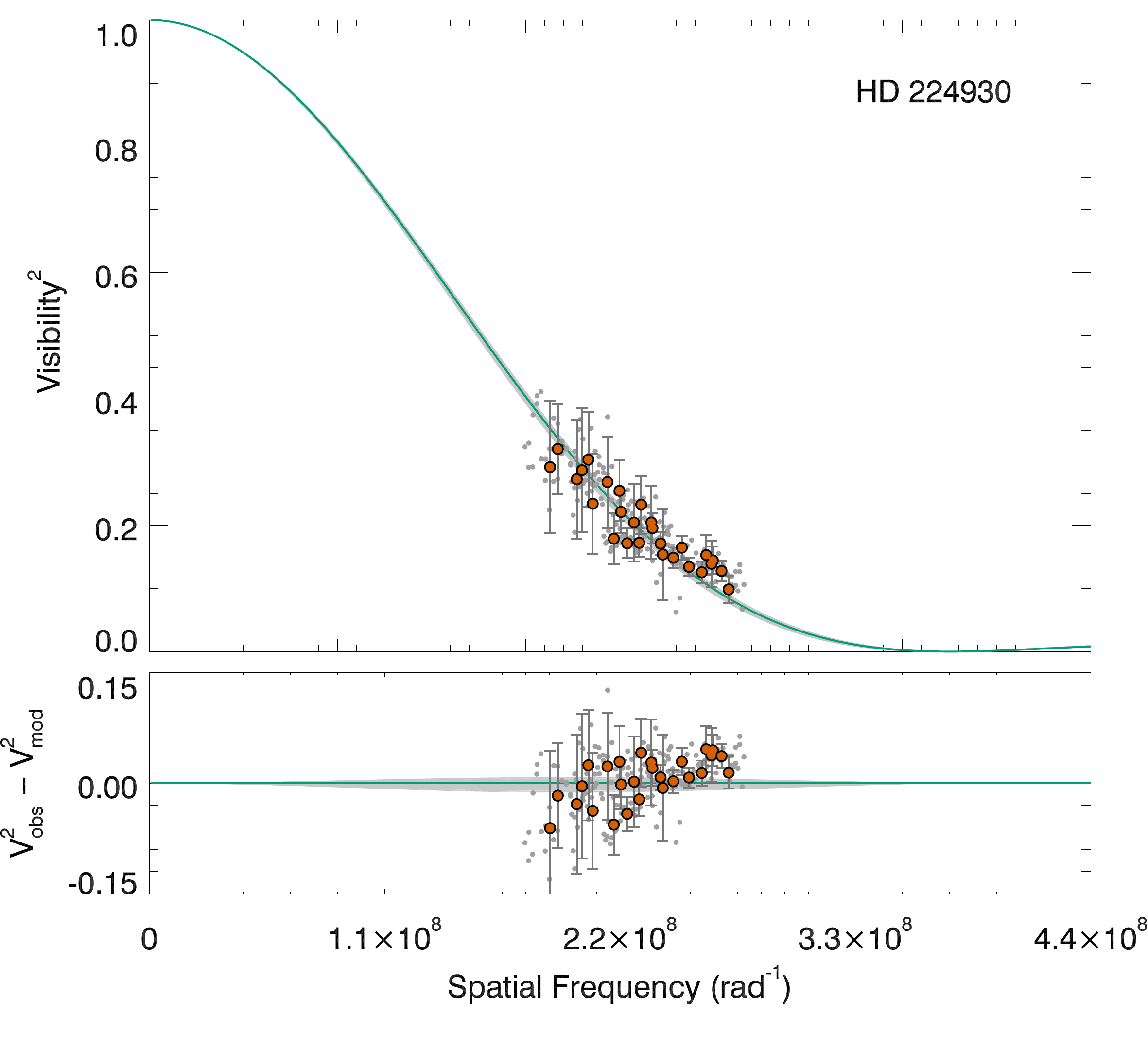}

            \caption{Squared visibility versus spatial frequency for HD\,221170 and HD\,224930.
                         The lower
      panel shows the residuals from the fit.  
            The error bars have been scaled to the reduced $\chi^{2}$. For HD\,221170 the reduced $\chi^{2}$\,=\,2.1 and for HD\,224930 $\chi^{2}$\,=\,8.4.
       All lines and symbols are the same as for Fig.\ref{figures1}.
      }

         \label{figures5}
   \end{figure}

\section{Methods and analysis}
\label{analysis}

In the following section we describe the method delivering the stellar
parameters, showing the connection between the interferometric,
photometric and spectroscopic analysis. To obtain the angular
diameter (see below), and hence the $T_\mathrm{eff}$, from the interferometric data requires
a limb-darkening parameter. This depends on $T_\mathrm{eff}$, $\log(g)$ and $[\mathrm{Fe/H}]$.
The process of estimating the $T_\mathrm{eff}$ is thus initiated by entering a
    first guess for the stellar parameters  (from the literature), and linearly
interpolating the limb-darkening coefficients from the STAGGER-grid \citep{magic15}.

The first limb-darkened angular diameter together with the bolometric
flux allows to directly compute the $T_\mathrm{eff}$ (equation~\ref{eq:teff}). The $\log(g)$ and $[\mathrm{Fe/H}]$ were then refined by isochrone fitting and spectroscopic
analysis: $\log(g)$ is sensitive to $T_\mathrm{eff}$ and metallicity, and $[\mathrm{Fe/H}]$ is
sensitive to $T_\mathrm{eff}$ and log(g), therefore, these values are slightly
refined with each iteration. The final values of fundamental stellar
parameters of the benchmark stars were iterated between
interferometric, photometric, and spectroscopic modelling, until
convergence was reached.

We did not encounter any major convergence problems. Changing the
initial guess parameters by 500\,K in $T_\mathrm{eff}$,
0.2\,dex in log(g), or 0.2\,dex in [Fe/H] 
did not change the final converged angular diameter result
(to within the $1\sigma$  errors).

\subsection{Modelling of limb-darkened angular diameters}
\label{models}

The determination of accurate angular diameters requires an estimate of an
appropriate amount of limb-darkening derived from stellar model atmospheres. As
a first step, we fitted to the visibility curves an undarkened uniform disc.
For all our fits, both with and without limb-darkening, we used a
least-squares fitting routine in IDL \citep[MPFIT,][]{markwardt09}, with uncertainties being determined by Monte Carlo simulations that took into account the uncertainty in the visibility measurements, as well as the wavelength calibration (5\,nm), calibrator sizes (5\%) and, for the limb-darkened fits, the limb-darkening coefficients.

Our fitted uniform disk diameters are listed in Table~\ref{angulardiam}. We also
fitted the commonly used linear limb-darkening law from \citet{claret11}; these are grids of coefficients calculated for various
model atmospheres and different photometric filters. For reference,
we also present the resulting limb-darkened
angular diameters in Table~\ref{angulardiam}. However, we stress that our final $T_\mathrm{eff}$
estimates are based on high-order limb-darkening coefficients from the
STAGGER-grid.
The 3D hydrodynamical models have been shown to better reproduce the solar
limb darkening than both theoretical and semi-empirical 1D hydrostatic models \citep{pereira13}. For this reason, they are expected to give better overall results and are adopted in the present analysis. The final results based on the STAGGER-grid are presented in Table~\ref{tab:observed}, and discussed below.

For robust estimates and accurate angular diameter we
employed higher-order limb-darkening laws.
The method used in this study
generally follows the same procedure 
described in Section 2.2
in the previous study of the same topic in \citet{karovicova18}.
In short, we employed the four-parameter 
limb-darkening coefficients of \citet{magic15}, 
that were calculated from 3D synthetic spectra from \citet{chiavassa18} for the STAGGER-grid of ab initio 3D hydrodynamic stellar atmosphere simulations \citep{magic13}.
These coefficients are tabulated as functions of 
$T_\mathrm{eff}$, $\log(g)$ and $[\mathrm{Fe/H}]$;
we interpolated them based on our initial guesses, and 
refined them using our measurements of $T_\mathrm{eff}$ based on the 
bolometric flux (Section~\ref{bolflux}), $\log(g)$ based on stellar evolution 
models (Section~\ref{stellarmodels}), and $[\mathrm{Fe/H}]$ based on spectroscopy 
(Section~\ref{spectroscopicanalysis}). 
We note that for one of our stars, HD\,221170, its $\log(g)$ value places it outside the STAGGER-grid. For this star we therefore linearly extrapolated its coefficients from the STAGGER-grid, and confirmed that these provided reasonable values by comparing them with coefficients from the tables of \citet{claret11}.
Using 3D models instead of 1D models generally has a very small effect on the
determined limb darkened angular diameters, compared to the error bars,
indicating that the measurements are usually only mildly dependent on the model
assumptions. However, in the worst case (HD\,122563) the  differences are 2\%,
that translates into 1\% in $T_\mathrm{eff}$ which is the targeted precision.
We present the limb darkening coefficients from the STAGGER-grid (in all 38
channels) in Tables 9-18 available at the CDS.

\subsection{Bolometric flux}
\label{bolflux}

Many of the stars in the sample have saturated or unreliable 2MASS photometry,
which prevents us from using the InfraRed Flux Method to derive bolometric
fluxes \citep{Casagrande10}. Hence, for all targets we use bolometric
corrections from \cite{Casagrande_VandenBerg14,Casagrande_VandenBerg18a}. We
use Hipparcos $H_p$ and Tycho2 $B_TV_T$ magnitudes for all stars, and 2MASS
$JHK_S$ only if with quality flag `A'. We assumed no reddening for all stars closer than 100 pc; for stars further away we estimated $E(B-V)$ using interstellar Na I D lines when possible, or the \cite{green15} map otherwise. 

Tables of bolometric corrections\footnote{\url{https://github.com/casaluca/bolometric-corrections}} were interpolated at the adopted reddening, and spectroscopic $[\mathrm{Fe/H}]$ and $\log(g)$. Spectroscopic $T_\mathrm{eff}$ were used only as a starting point to interpolate bolometric corrections. 
The adopted bolometric corrections are listed  in Table~\ref{bolcor}.
An iterative procedure was adopted, where the bolometric corrections were used
together with the angular diameter to derive an updated $T_\mathrm{eff}$  until
convergence was reached to within a few K.

The bolometric flux was obtained
using a weighted average of the bolometric flux from the bolometric correction
in each band. Weights were given by the inverse of the estimated variance
of the bolometric flux derived from each band. These were obtained for each
photometric band by computing the mean square deviation using a Monte Carlo
integration over four independent parameters ($T_\mathrm{eff}$, $\log(g)$,
$[\mathrm{Fe/H}]$ and $E(B-V)$) and the photometric magnitude for that band.
All five input parameters errors were modelled as independent normally
distributed random variables. The uncertainties quoted for the bolometric flux
are the square root of the weighted sample variance, plus a 0.3\% systematic to
account for the uncertainty in the adopted reference solar luminosity. The
systematic uncertainties and inaccuracies stemming from the use of model fluxes
are harder to quantify, but extensive comparison with absolute
spectrophotometry in \citet{casagrande18} indicates that bolometric fluxes are
typically recovered at the percent level for FG stars. Our sample comprises
cooler stars, for which the performances of our bolometric corrections are much
less tested. Reassuringly, the comparison of our bolometric corrections with
absolute spectrophotometry from \citet{white18} also indicates good agreement
for stars in the $T_\mathrm{eff}$ range covered by the present work.

{
\begin{sidewaystable*}
   \caption{Bolometric corrections}    
  \centering   
  \begin{tabular}{l l l l l l l l l l l l l l l l l l l}      
\hline\hline   
 Star        &     BC$_{H_p}$ &  BC$_{B_T}$ &  BC$_{V_T}$ &  BC$_J$ &  BC$_H$
      &  BC$_K$ &   $B_T$  &    e$B_T$    &  $V_T$   &  e$V_T$   &          $H_p$     & e$H_p$   &     $J$  &    e$J$  &    $H$  &    e$H$  &    $K$  &   e$K$ \\
 & & & & & & & & & & & & & & & & & & \\
 \hline
HD\,2665     &    -0.622 & -1.494 & -0.592 & 1.306 &  1.827 & 1.949 &  8.649&	0.016 &	7.813 &	0.010  &	  7.8637 & 0.0009&   6.026&  0.024 &  5.603&  0.033 &  5.474&  0.016 \\
 & & & & & & & & & & & & & & & & & & \\
HD\,6755     &    -0.489 & -1.324 & -0.453 & 1.334 &  1.842 & 1.955 &  8.586&	0.016 &	7.808 &	0.010  &	  7.8620 & 0.0015&   6.197&  0.024 &  5.759&  0.020 &  5.662&  0.017 \\
 & & & & & & & & & & & & & & & & & & \\
HD\,6833     &    -0.828 & -2.135 & -0.814 & 1.455 &  2.107 & 2.256 &  8.245&	0.015 &	6.892 &	0.009  &	  6.9110 & 0.0007&     -  &    -   &    -  &    -   &    -  &    - \\
 & & & & & & & & & & & & & & & & & & \\
HD\,103095   &    -0.357 & -1.105 & -0.318 & 1.258 &  1.682 & 1.786 &  7.353&	0.015 &	6.509 &	0.010  &	  6.5641 & 0.0009&     -  &    -   &    -  &    -   &  4.373&  0.027 \\
 & & & & & & & & & & & & & & & & & & \\
HD\,122563   &    -0.603 & -1.597 & -0.578 & 1.438 &  2.008 & 2.116 &  7.275&	0.015 &	6.304 &	0.009  &	  6.3275 & 0.0007&     -  &    -   &    -  &    -   &    -  &    - \\
 & & & & & & & & & & & & & & & & & & \\
HD\,127243   &    -0.415 & -1.314 & -0.369 & 1.345 &  1.824 & 1.931 &  6.655&	0.014 &	5.681 &	0.009  &	  5.7362 & 0.0005&     -  &    -   &    -  &    -   &    -  &    - \\
 & & & & & & & & & & & & & & & & & & \\
HD\,140283   &    -0.278 & -0.739 & -0.244 & 1.017 &  1.345 & 1.421 &  7.771&	0.016 &	7.269 &	0.011  &	  7.3000 & 0.0013&   6.014&  0.019 &  5.696&  0.036 &  5.588&  0.017 \\
 & & & & & & & & & & & & & & & & & & \\
HD\,175305   &    -0.501 & -1.372 & -0.465 & 1.350 &  1.871 & 1.983 &  8.109&	0.015 &	7.275 &	0.010  &	  7.3211 & 0.0006&   5.613&  0.023 &  5.167&  0.024 &  5.057&  0.020 \\
 & & & & & & & & & & & & & & & & & & \\
HD\,221170   &    -1.050 & -2.501 & -1.062 & 1.495 &  2.204 & 2.368 &  8.990&	0.018 &	7.797 &	0.011  &	  7.8373 & 0.0009&   5.532&  0.018 &  4.987&  0.044 &  4.836&  0.016 \\
 & & & & & & & & & & & & & & & & & & \\
HD\,224930   &    -0.281 & -0.967 & -0.234 & 1.186 &  1.560 & 1.648 &  6.555&	0.014 &	5.834 &	0.009  &	  5.8735 & 0.0009&     -  &    -   &    -  &    -   &    -  &    - \\
   \hline  
\end{tabular}
\tablefoot{Adopted bolometric corrections (BC). Hipparcos $H_p$ and Tycho2
       $B_TV_T$ magnitudes for all stars. 2MASS $JHK_S$ only if the quality is
       flag `A' The zero-point of these bolometric corrections is set by $M_{bol,\odot}=4.75$.}
\label{bolcor}
  \end{sidewaystable*}
}

\subsection{Stellar evolution models}
\label{stellarmodels}

We used the ELLI package\footnote{Available online at \url{https://github.com/dotbot2000/elli}} \citep{Lin18} to estimate stellar masses based on comparisons to Dartmouth stellar evolution tracks \citep{Dotter2008}, computed with alpha enhancement. The comparison uses a Bayesian framework to estimate the stellar mass and age from $T_\mathrm{eff}$, $\log L/L_\odot$ and $[{\rm Fe }/{\rm H}]$, taking into account their related (assumed independent) errors. An initial guess is produced from a maximum likelihood estimate at our estimated metallicity, between the fundamental stellar parameters and those estimated on the isochrone. A Markov chain Monte Carlo (MCMC) method is then used to sample the posterior distribution, and we take the mean and dispersion on this distribution as our estimate for the mass and its uncertainty.
Finally, we compute the surface gravity from its fundamental relation, rewritten to a form that directly utilises the measurements, 
\begin{equation}
\log g = \log \frac {G M}{R^2} = \log \frac {4 G M \varpi^2}{\theta^2},
\label{eq:logg}
\end{equation}
where $G$ is the gravitational constant and $\varpi$ the parallax.

As shown in Fig~\ref{kiel}, there are systematic offsets between the theoretical stellar isochrones and the parameters of metal-poor stars on the red giant branch. Our Bayesian sampling approach therefore does a poor job of predicting the properties of these stars. Instead, we adopted the turnoff-mass at the relevant metallicity and assuming an age $>10$\,Gyr. 
Since we did not use the Bayesian approach for these stars, we use instead a conservative uncertainty estimate on the stellar mass of 0.05\,$M_\odot$.

\subsection{Spectroscopic analysis}
\label{spectroscopicanalysis}

High-resolution spectra for the stars were extracted from the 
ELODIE \citep[$R \approx 42\,000$,][]{Moultaka2004} and 
FIES \citep[$R \approx 65\,000$,][]{Telting2014} archives. 
We determined the stellar iron abundances using a custom pipeline based on the spectrum synthesis code SME \citep{piskunov_spectroscopy_2017}
using MARCS model atmospheres \citep{gustafsson_grid_2008}
and pre-computed non-LTE departure coefficients for Fe \citep{amarsi_non-lte_2016-1}.\\

We selected unblended lines of \ion{Fe}i and \ion{Fe}{ii} between 4400 and 6800\,\AA\ with 
accurately known oscillator strengths from laboratory measurements. 
For saturated lines, we ensured that collisional broadening parameters were available from ABO theory \citep{barklem_list_2000,barklem_broadening_2005}.
To obtain a differential $[\mathrm{Fe/H}]$, solar abundances were also measured from solar spectra recorded with the same spectrographs as our target stars, based on observations of light reflected off the Moon (ELODIE) and Vesta (FIES). We thereby produce solar-differential abundances, which mostly cancels uncertainties in oscillator strengths as well as potential systematic differences between the spectrographs.
We estimated the iron abundance of each star from the
outlier-resistant mean of the entire set of \ion{Fe}{I} and
\ion{Fe}{II} lines, with $3 \sigma$ clipping.
We also computed the difference in abundance between lines of \ion{Fe}i and \ion{Fe}{ii}, as an estimate of how closely our fundamental stellar parameters fulfill the ionization equilibrium.
Finally, we compute a systematic uncertainty on the metallicity, that we derive by perturbing the input parameters one at a time according to their formal errors, and add these differences in quadrature.

   \begin{table*}
   \caption{Derived stellar parameters ($T_\mathrm{eff}$ , $[\mathrm{Fe/H}]$, $\log(g)$)}    
  \centering   
  \begin{tabular}{l l c c }      
\hline\hline   
 
 Star &   \multicolumn{1}{c}{$T_\mathrm{eff}$}  &    $\log(g)$ &     $[\mathrm{Fe/H}]$\tablefootmark{a}  \\
    &      \multicolumn{1}{c}{(K)} & (dex)& (dex) \\ 
  \hline 
   HD\,2665    & 4883 $\pm$ 95  & 2.209 $\pm$ 0.032 & $-$2.10 $\pm$ 0.09 $\pm$ 0.10  \\
   HD\,6755    & 4888 $\pm$ 131 & 2.685 $\pm$ 0.031 & $-$1.71 $\pm$ 0.10 $\pm$ 0.14  \\
   HD\,6833    & 4438 $\pm$ 141 & 1.860 $\pm$ 0.072 & $-$0.80 $\pm$ 0.07 $\pm$ 0.04  \\
   HD\,103095  & 5174 $\pm$ 32  & 4.702 $\pm$ 0.015 & $-$1.26 $\pm$ 0.07 $\pm$ 0.02  \\
   HD\,122563  & 4635 $\pm$ 34  & 1.404 $\pm$ 0.035 & $-$2.75 $\pm$ 0.12 $\pm$ 0.04  \\
   HD\,127243  & 4959 $\pm$ 21  & 2.599	$\pm$ 0.047 & $-$0.71 $\pm$ 0.06 $\pm$ 0.02  \\
   HD\,140283  & 5792 $\pm$ 55  & 3.653 $\pm$ 0.024 & $-$2.29 $\pm$ 0.10 $\pm$ 0.04  \\
   HD\,175305  & 4850 $\pm$ 118 & 2.502 $\pm$ 0.031 & $-$1.52 $\pm$ 0.08 $\pm$ 0.12  \\
   HD\,221170  & 4248 $\pm$ 128 & 1.251 $\pm$ 0.042 & $-$2.40 $\pm$ 0.13 $\pm$ 0.17  \\
   HD\,224930  & 5422 $\pm$ 28  & 4.337 $\pm$ 0.012 & $-$0.81 $\pm$ 0.05 $\pm$ 0.02  \\
   \hline  
\end{tabular}
\tablefoot{\tablefoottext{a}{The error bars on $[\mathrm{Fe/H}]$ denote the statistical measurement uncertainty, and the systematic error propagated from $T_\mathrm{eff}$ and $\log(g)$, respectively.}}
\label{finalparameters}
  \end{table*}

   \begin{table}
       \caption{{Uncertainties in $T_\mathrm{eff}$ and how they propagate
       from the underlying measurements}}
  \centering   
  \begin{tabular}{l l r r r }      
\hline\hline   
 
 Star &   $T_\mathrm{eff}$  & e$T_\mathrm{eff}$   & e$F_\mathrm{bol}$ $^{a}$ & e$\varTheta_{LD}$ $^{b}$ \\
    &      (K) & (K)& (K) & (K)\\

  \hline 
   HD\,2665    & 4883    & 95   & \bf{92}   & 25      \\
   HD\,6755    & 4888    & 131  & \bf{128}  & 26      \\
   HD\,6833    & 4438    & 141  & \bf{139}  & 21      \\
   HD\,103095  & 5174    & 32   & \bf{27}   & 17      \\
   HD\,122563  & 4635    & 34   & 19        & \bf{28}  \\
   HD\,127243  & 4959    & 21   & 12        & \bf{18} \\
   HD\,140283  & 5792    & 55   & 11        & \bf{54} \\ 
   HD\,175305  & 4850    & 118  & \bf{114}  & 30      \\
   HD\,221170  & 4248    & 128  & \bf{126}  & 18      \\
   HD\,224930  & 5422    & 28   & 9         & \bf{27} \\
   \hline  
\end{tabular}

\tablefoot{\tablefoottext{a}{The uncertainties contribution from the bolometric flux if the
$\varTheta_{LD}$ uncertainties are set to 0.
$^{(b)}$  The uncertainties arising entirely from the angular diameter measurements if the $F_\mathrm{bol}$
 uncertainties are set to 0.}}
\label{errors}
  \end{table}

\section{Results and discussion}
\label{results}
\subsection{Recommended stellar parameters}

We presented fundamental stellar parameters and 
angular diameters for a set of benchmark stars.
Four of the ten stars are 
{\it Gaia} FGK benchmark stars (HD\,12256, HD\,103095, HD\,140283, HD\,175305).
Six of the stars (HD\,2665, HD\,6755, HD\,6833, HD\,221170, 
HD\,127243, HD\,224930) were added and suggested
as new benchmark stars.
For all stars we estimated 
$T_\mathrm{eff}$, log$g$, $[\mathrm{Fe/H}]$ and $\theta$$_{LD}$.
All the values along with mass, luminosity 
and radii are summarized in Table~\ref{tab:observed}.

\subsection{Uncertainties}
\label{uncertainties}

The final $T_\mathrm{eff}$ uncertainties 
consist of uncertainties in the bolometric flux and 
the uncertainties in the angular diameter.
Table~\ref{errors} shows the contribution of each part.
The third column shows the final $T_\mathrm{eff}$ uncertainties,
the fourth column the uncertainties raising from the bolometric
flux if the $\varTheta_{LD}$ uncertainties are set to 0.
The fifth column shows the $F_\mathrm{bol}$ uncertainties 
set to 0, with the uncertainties raising entirely from the
angular diameter.

The statistical measurement uncertainties in 
$\log(g)$ and $\mathrm{[Fe/H]}$ from the
isochrone fitting and spectroscopic analysis were folded into the uncertainties
of the angular diameters and thus are included in the final 
$T_{\mathrm{eff}}$ error
estimates. The median uncertainties in 
$\log(g)$ and $\mathrm{[Fe/H]}$ across our sample of
stars are 0.03~dex and 0.09~dex, respectively (Table~\ref{finalparameters}).

For five of the stars, the final $T_\mathrm{eff}$ uncertainties are less than
around 50~K, or 1\%.  For these stars, the errors coming from the
bolometric flux are less than or similar to those coming from the
limb-darkened angular diameter. The final $T_\mathrm{eff}$ uncertainties for the
other five stars are somewhat larger: 100--150~K. This is driven by
larger errors in the bolometric flux, rather than in the angular
diameter. As mentioned above, the precision that is desired by the
spectroscopic
teams of surveys like {\it Gaia}-ESO or GALAH is around 1\% (or around 40-60~K); we achieve this for half of our sample, and could achieve it for the
full sample if more precise bolometric fluxes are available.

\subsection{Comparison with literature values}

Three of our ten targets (HD\,103095,
HD\,122563 and HD\,140283)
were previously interferometrically studied by \citet{Creevey12, Creevey15}
and they are also a part of the previous interferometric study
\citep{karovicova18}.
These stars were used as {\it Gaia} FGK benchmark stars
in the {\it Gaia}-ESO spectroscopic survey. However, the stars HD\,103095,
and HD\,140283 had to be
reconsidered as their $T_\mathrm{eff}$
did not reconcile with spectroscopic studies.
Therefore, the stars were not recommended as 
temperature standards until discrepancies are resolved \citep[see][]{Heiter15}.
 The issues were resolved in \citet{karovicova18} and the stars can be now again used as benchmarks.\\

{\bf HD\,103095} This star was interferometrically observed by \citet{Creevey12}, who reported
$T_\mathrm{eff}$=4818$\pm$54~K. This value is lower than a value estimated in
the previous study \citep{karovicova18}
where $T_\mathrm{eff}$=5140$\pm$49~K was determined.
Here with our improved reduction method we obtain
$T_\mathrm{eff}$=5174$\pm$32, log$g$=4.702$\pm$0.015~dex and
$[\mathrm{Fe/H}]$=$-$1.26$\pm$0.07~dex; note all the differences with the previous study are within the stipulated uncertainties.\\

{\bf HD\,122563} This metal-poor star is well studied spectroscopically. It
 was included in the {\it Gaia} FGK benchmark sample with $T_\mathrm{eff}$=4587$\pm$60~K and log$g$=1.61$\pm$0.07~dex {\citep{Heiter15}}. The star was also a part of the interferometric study by \citet{Creevey12}. The reported 
$T_\mathrm{eff}$=4598$\pm$41~K by \citet{Creevey12} agrees within the uncertainties with our estimated value.
The $T_\mathrm{eff}$ value from \citet{karovicova18} is $T_\mathrm{eff}$=4636$\pm$37~K,
the updated value is $T_\mathrm{eff}$=4635$\pm$34 together with
log$g$=1.404$\pm$0.035~dex and $[\mathrm{Fe/H}]$=$-$2.75$\pm$0.12~dex. The
$T_\mathrm{eff}$ is in agreement with expected photometric and spectroscopic value.\\

{\bf HD\,140283} This very metal-poor star was interferometrically measured by \citet{Creevey15}. There were two $T_\mathrm{eff}$ reported based on two different reddening
and $T_\mathrm{eff}$ is thus between $T_\mathrm{eff}$=5534$\pm$103~K and 5647$\pm$105~K. 
These values were in disagreement with spectroscopy and
photometry.
 The $T_\mathrm{eff}$=5787$\pm$48~K determined in \citet{karovicova18} was in comparison to \citet{Creevey15}
253\,K and 140\,K higher, respectively,
bringing the interferometric values also into disagreement. 
 The issues were resolved and put the spectroscopic, photometric and interferometric values into better agreement.
The new $T_\mathrm{eff}$=5792$\pm$55 and other stellar parameters are: log$g$=3.653$\pm$0.024~dex and $[\mathrm{Fe/H}]$=$-$2.29$\pm$0.10~dex.
The differences between the interferometrically determined $T_\mathrm{eff}$ 
of \citet{Creevey12, Creevey15} and \citet{karovicova18}
are result of differences in measured angular diameters 
of the stars. This points to systematic
errors arising from the known difficult calibration of
interferometric observations, especially of the smaller targets.\\

The rest of the stars were previously not interferometrically studied, however, for comparison we list various spectroscopic parameters
as published in the PASTEL catalogue \citep{pastel10}.
Our values of $T_\mathrm{eff}$, log$g$ and $[\mathrm{Fe/H}]$ 
are listed in Table~\ref{finalparameters}.
We compare our values with spectroscopical studies
executed after 2000 when high resolution spectroscopic instruments were available.
For details on uncertainties of our values please look Table~\ref{finalparameters} and ~\ref{errors} as well as section \ref{uncertainties}.\\

{\bf HD\,175305} \citet{hawkins16} suggested this star as a benchmark.
They derived stellar parameters for it
by averaging different values from the PASTEL catalogue, and arrived at
 $T_\mathrm{eff}$=5085$\pm$58~K, log$g$=2.49$\pm$0.25~dex and $[\mathrm{Fe/H}]$=-1.43$\pm$0.07~dex \citep[see][]{hawkins16}.
The stellar parameters were compiled using the PASTEL database \citep{soubiran10}. 
We report $T_\mathrm{eff}$=4850$\pm$118~K, log$g$=2.502$\pm$0.031~dex and $[\mathrm{Fe/H}]$=$-$1.52$\pm$0.08~dex.
Our values point to a much cooler star.\\

{\bf HD\,2665} According to the PASTEL catalogue the $T_\mathrm{eff}$ measurements range between 5000-5123~K, with log$g$ between 2.20-2.35
and metallicity of -1.9. Our $T_\mathrm{eff}$ of 4883$\pm$95~K is significantly lower. The log$g$ within the range and
with lower metallicity of -2.1$\pm$0.09~dex.\\

{\bf HD\,6755} In the PASTEL catalogue the $T_\mathrm{eff}$ measurements
range between 5011-5169~K, with only two values for each log$g$ 2.7 and 2.8~dex and for $[\mathrm{Fe/H}]$ -1.47 and -1.58~dex. Our value for $T_\mathrm{eff}$ is again systematically much lower, 4888$\pm$131~K, however with a rather large uncertainty of 131~K arising from the bolometric flux estimate. 
We also determine a slightly lower metallicity
of -1.7$\pm$0.10~dex in comparison to the literature values.\\

{\bf HD\,6833} The PASTEL catalogue shows only two values for this star
$T_\mathrm{eff}$ of 4400 and 4450~K, log$g$ of 1 and 1.4~dex and 
$[\mathrm{Fe/H}]$ of -0.89 and -1.04~dex.
Our values agree, $T_\mathrm{eff}$=4438$\pm$141~K and $[\mathrm{Fe/H}]$=-0.8$\pm$0.07~dex. However, we present higher 
log$g$=1.860~dex$\pm$0.072.\\

{\bf HD\,127243} According to the PASTEL catalogue, this subgiant has been studied spectroscopically 
4 times, with $T_\mathrm{eff}$ measurements ranging between 5000 and 5350~K, 
surface gravity between 2.2 and 3.5 and metallicity between -0.6 and -0.7. 
Our estimate of the $T_\mathrm{eff}$ shows a value close to the lower range (4959$\pm$21~K),
while other stellar values are within the range.\\

{\bf HD\,221170} The literature values from the PASTEL
catalogue show a slightly warmer star
with higher metallicity than our estimated values.
$T_\mathrm{eff}$ between 4425-4648~K, 
log$g$ between 0.9-1.05~dex and $[\mathrm{Fe/H}]$ between -2 and -2.190~dex.
Our temperature is significantly lower, with
$T_\mathrm{eff}$ of 4248$\pm$128~K. We present log$g$ of 1.251$\pm$0.042~dex and our results also
show the star to be more metal poor with
$[\mathrm{Fe/H}]$ of -2.4$\pm$0.13~dex.\\

{\bf HD\,224930} According to the PASTEL catalogue, this star has been studied spectroscopically several 
times and the reported $T_\mathrm{eff}$
is widely spread between 5169~K and 5680~K, log$g$ between 
4.1 and 4.5~dex and $[\mathrm{Fe/H}]$ between -0.52 and -1.
Our values lie in the middle of the spread with
$T_\mathrm{eff}$ of 5422$\pm$28~K, log$g$ of 4.337$\pm$0.012~dex and 
$[\mathrm{Fe/H}]$ of -0.81$\pm$0.05~dex.

\subsection{Fe ionization balance}
 The relative populations of different ionization stages is a sensitive measure of atmospheric properties. The so-called ionization balance involves
 matching the overall Fe elemental abundance as derived from \ion{Fe}{i} and
 \ion{Fe}{ii} in order to determine a star's surface gravity
 {\citep{tsantaki19}.} Conversely, when the surface gravity is already known the ionization balance can instead be used to infer an effective temperature \citep[see, e.g.,][]{Bergemann12}, or to verify the consistency of the two.

\begin{figure}
    \includegraphics[width=\hsize]{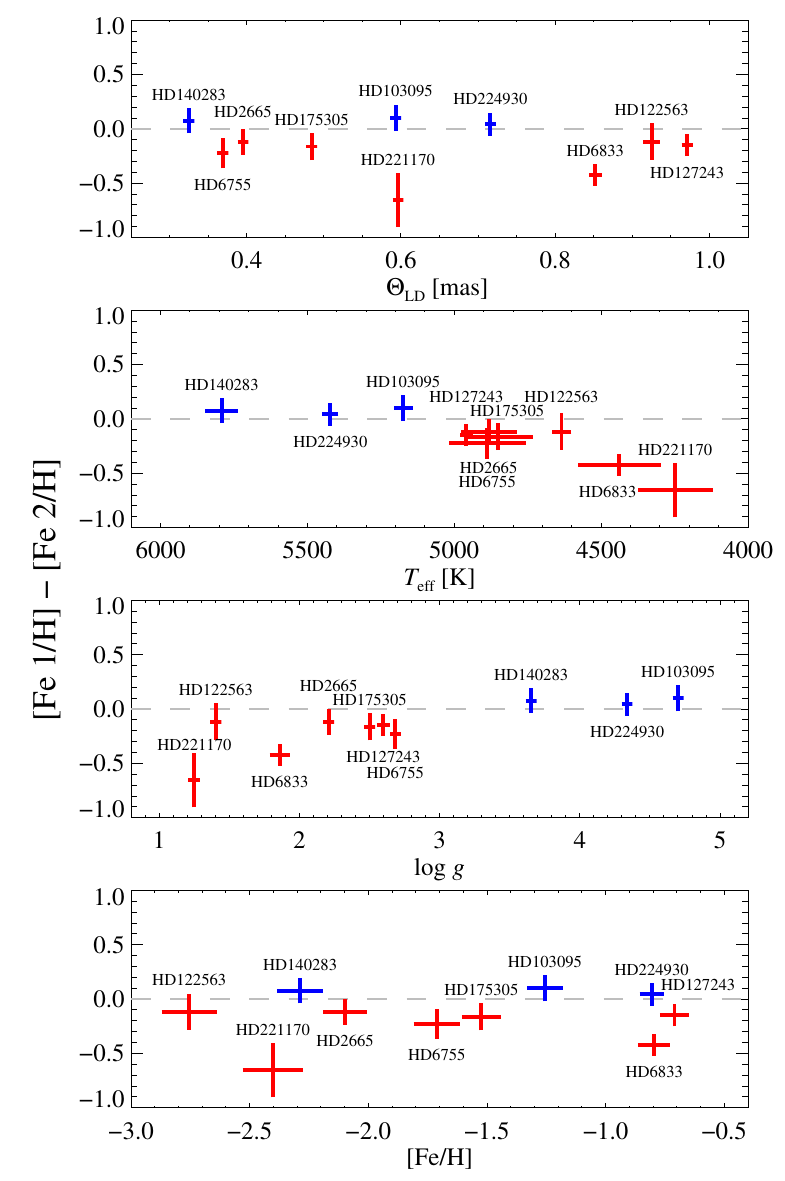}
\caption{{Deviations from ionization balance, i.e. the difference between the
    abundances determined from lines of neutral and ionized iron, as a function
    of the measured stellar parameters. Vertical and horizontal lines represent
    the combined uncertainties from the two measurements.  Each star is
    labelled, and colour-coded as red for red giants or blue for main-sequence and subgiants. 
    }}
\label{fig:ionisation_equilibrium}
\end{figure}

We find that our iron abundance determinations generally yield acceptable agreement for lines of neutral and ionized iron. We illustrate in Fig.~\ref{fig:ionisation_equilibrium} these abundance differences as a function of the measured angular diameters and stellar parameters. 
The abundance differences are small for the dwarf stars in the sample, consistent with their statistical uncertainties. Among the giant stars however, we find a strong trend with $T_\mathrm{eff}$ such that the coolest stars deviate strongly from ionization equilibrium by upwards of 0.5~dex. However, we find that these discrepancies do not correlate with angular diameters, indicating that they are not driven by instrumental artefacts but rather by shortcomings in the spectroscopic analysis.
We do however identify a trend between the abundance differences and the 
effective temperature, where the coolest stars in our sample show increasingly
large deviations from ionization balance exceeding 0.4~dex for HD\,6833 (4438\,K) and 0.6\,dex for HD\,221170 (4248\,K). 

Importantly, this indicates that a non-differential spectroscopic derivation of stellar parameters for cool, very metal-poor stars cannot accurately recover their surface gravity.
3D non-LTE models could help to resolve this discrepancy \citep[e.g.][]{amarsi16,amarsi19}.

The measurement of iron abundances from lines of the neutral species is sensitive to the adopted effective temperature, where a change of $\pm100$\,K will on average affect the measured abundance by $\pm0.07$\,dex. The corresponding effect on lines of ionized iron is of the order $\pm0.02$ and $\pm0.05$\,dex for stars warmer and cooler than 5500\,K, respectively. Conversely, a change in $\log(g)$ of $\pm0.1$\,dex will affect the abundance from lines of neutral iron by less than 0.01\,dex. For ionized lines, the corresponding effect on the abundance difference is $\pm0.05$\,dex. An error in $T_\mathrm{eff}$ of $\pm100$\,K will therefore typically affect the difference in abundances from lines of neutral iron relative to ionized, by of the order $\pm0.1$\,dex, and an error in $\log(g)$ of $\pm0.1$\,dex would have a corresponding effect of $\pm0.05$\,dex. Errors in [Fe/H] from \ion{Fe}{I} and from \ion{Fe}{II} could thereby partially cancel.

\section{Conclusions}

This project delivered fundamental stellar 
parameters for ten metal-poor stars.
Stars with low metallicity are poorly represented in
benchmark sample used so far. 
Reliable angular diameters for metal poor stars have been difficult to measure so far because these stars are faint for suitable interferometric instruments.
We took this into consideration, observed the stars over various 
nights, baseline configurations and tried to resolve the 
targets close to the first null of the visibility curve.
We observed the stars using the high angular resolution 
instrument PAVO and the CHARA array and we measured 
accurate angular diameters of the stars.

In order to estimate the limb darkening diameters,
we used the 3D radiation-hydrodynamical 
model atmospheres in the STAGGER-grid.
The $T_\mathrm{eff}$ were directly computed from the 
Stefan-Boltzmann relation using the
measured angular diameters and bolometric flux.
Bolometric fluxes were computed from multi-band photometry interpolating iteratively on a grid of synthetic stellar fluxes, to ensure consistency with the final adopted stellar parameters. 
High resolution spectroscopy
allowed us to determine $[\mathrm{Fe/H}]$, isochrone fitting to derive 
mass, and parallax measurements to constrain the absolute luminosity. 
After iterative refinement we derived the final fundamental 
parameters of $T_\mathrm{eff}$, $\log(g)$, $[\mathrm{Fe/H}]$.

This allowed us to reach the  
desired precision of better than 1\% in the $T_\mathrm{eff}$
for 5 stars in our sample 
HD\,103095, HD\,122563, HD\,127243,
HD\,140283 and HD\,224930.
A precision of 1\% in 
$T_\mathrm{eff}$ is essential for the correct determination of the atmospheric parameters of the survey sources.
For the remaining stars, for which the uncertainties in the $T_\mathrm{eff}$
are higher than 1\%,
the uncertainty in the
bolometric flux significantly contributes to
the final uncertainty in the effective 
temperature ($\sim$2-3\%).
For all stars in our sample
we determined $\log(g)$  
and $\mathrm{[Fe/H]}$, with median uncertainties of 
0.03~dex and 0.09~dex, respectively.

We presented the first from the series of papers that are aiming to build a new robust sample of benchmark stars. The reliable interferometric stellar parameters presented here should be useful for testing and validating stellar analysis pipelines {\citep{jofre19}}, that typically rely on photometric and spectroscopic methods.
Our consistent measurements and analysis will also help to cross-calibrate different large stellar surveys such as {\it Gaia} \citep{gaia18}, APOGEE \citep{alendeprieto08}, {\it Gaia}-ESO Survey \citep{Gilmore12, Randich13}, 4MOST \citep{dejong12}, WEAVE \citep{dalton12}, GALAH \citep{desilva15}. In turn, achieving these goals will help us to more robustly
understand 
the physics of stars, and uncover the structure and evolution of our Galaxy.

\begin{acknowledgements}
      I.K. acknowledges the German
      \emph{Deut\-sche For\-schungs\-ge\-mein\-schaft, DFG\/} project
      number KA4055 and by the European Science Foundation - GREAT {\it Gaia} Research for European Astronomy Training.
      M.I. was the recipient of an Australian Research Council Future Fellowship (FT130100235) funded by the Australian Government.
      P.J. acknowledges FONDECYT Iniciaci\'{o}n programme number 11170174.
      This work is based upon observations obtained with the Georgia State University Center for High Angular Resolution Astronomy Array at Mount Wilson Observatory. The CHARA Array is supported by the National Science Foundation under Grants No. AST-1211929 and AST-1411654. Institutional support has been provided from the GSU College of Arts and Sciences and the GSU Office of the Vice President for Research and Economic Development.
      This work is based on spectral data retrieved from the ELODIE archive at Observatoire de Haute-Provence (OHP), and on observations made with the Nordic Optical Telescope, operated by the Nordic Optical Telescope Scientific Association at the Observatorio del Roque de los Muchachos, La Palma, Spain, of the Instituto de Astrofisica de Canarias.
      Thanks to Prof. Gilmore for supporting observing and grant proposals through the whole project.
      Thanks to Dr. Th\'{e}venin for providing helpful comments and for his support of the project. Thanks to Dr. Creevey for her collaboration.
      Thanks to Dr. Lind for helpful discussions and for providing 
      preliminary spectroscopic computations.
      Finally,
      we are extremely grateful to the anonymous referee for carefully
      reading the manuscript, 
      and providing helpful comments.

\end{acknowledgements}

\bibliographystyle{aa} 
\bibliography{ref}

\end{document}